\documentclass{emulateapj}
\usepackage{apjfonts}

\begin{document}

\newcommand{\vdag}{(v)^\dagger}
\newcommand{\oii}{[O{\sc ii}] }
\newcommand{\myemail}{poggianti@pd.astro.it}
\newcommand{\sigss}{${\Sigma}_{2D}$}
\newcommand{\sigsss}{${\Sigma}_{3D}$}
\newcommand{\pog}{Poggianti et al. (2006) }


\slugcomment{Draft}
\slugcomment{\today}


\shorttitle{Environments of starbursts and post-starbursts at high-z}
\shortauthors{Poggianti et al.}




\title{The environments of starburst and post-starburst galaxies at z=0.4-0.8
}

\author{Bianca M. Poggianti$^1$, Alfonso Arag\'on-Salamanca, Dennis Zaritsky,
Gabriella De Lucia, Bo Milvang-Jensen, 
Vandana Desai, Pascale Jablonka, Claire Halliday, Gregory
Rudnick, Jesus Varela, Steven Bamford, Philip Best,
Douglas Clowe, Stefan Noll, Roberto Saglia, Roser
Pello, Luc Simard, Anja von der Linden, Simon White}

\email{bianca.poggianti@oapd.inaf.it}



\begin{abstract}
Post-starburst (E+A or k+a) spectra, characterized by their
exceptionally strong Balmer lines in absorption and the lack of
emission lines, belong to galaxies in which the star formation
activity ended abruptly sometime during the past Gyr.  We perform
a spectral analysis of galaxies in clusters, groups, poor groups and
the field at $z=0.4-0.8$ based on the ESO Distant Cluster Survey.  We
find that the incidence of k+a galaxies at these redshifts depends strongly
on environment. K+a's reside preferentially in clusters and,
unexpectedly, in a subset of the $\sigma = 200-400 \rm \, km \,
s^{-1}$ groups, those that have a low fraction of [OII] emitters. In these
environments, 20-30\% of the recently 
star-forming galaxies have had their star formation activity recently
truncated.  In contrast, there are proportionally fewer k+a galaxies
in the field, the poor groups and groups with a high [OII]
fraction. An important result is that the incidence of k+a galaxies
correlates with the cluster velocity dispersion: more massive clusters
have higher proportions of k+a's. 
Spectra of dusty starburst candidates, with strong Balmer absorption
and emission lines, present a very different environmental dependence
from k+a's. They are numerous in all environments at $z=0.4-0.8$, but
they are especially numerous in all types of groups, favoring the
hypothesis of triggering by a merger. We present the morphological
type, stellar mass, luminosity, mass-to-light ratio, local galaxy
density and clustercentric distance distributions of galaxies of
different spectral types. These properties are consistent with
previous suggestions that cluster k+a galaxies are observed in a
transition phase, at the moment they are rather massive S0 and Sa
galaxies, evolving from star-forming, recently infallen later types to
passively evolving cluster early-type galaxies. The correlation
between k+a fraction and cluster velocity dispersion supports
the hypothesis that k+a galaxies in clusters originate from processes
related to the intracluster medium, while several possibilities are
discussed for the origin of the puzzling k+a frequency in low-[OII]
groups.
\end{abstract}





\section{Introduction}

``Post-starburst'' galaxies, also known as ``E+A'' or ``k+a'' galaxies
\footnote{Sometimes referred to as ``HDS=Hdelta strong'' galaxies.},
are identified on the basis of their optical spectra for the
exceptionally strong Balmer lines in absorption and the lack of
emission lines (Dressler \& Gunn 1983, Couch \& Sharples 1987).

Extensive spectrophotometric modeling has demonstrated that this
spectral combination arises if the star formation (SF) activity
has ended abruptly sometime between $\sim 5 \times 10^7$ and
$\sim 1.5 \times 10^9$ years prior to observations (Couch \& Sharples
1987, Newberry, Boroson \& Kirshner 1990, Abraham et al. 1996, Barger
et al. 1996, Poggianti \& Barbaro 1996, 1997, Bekki, Shioya \& Couch
2001, Dressler et al. 2008, see Poggianti 2004 for a modeling review).
Only those spectra with the strongest Balmer lines require a recent
burst, while k+a spectra with moderate line strengths can be modelled
as post-starforming galaxies in which the truncation follows a
regular star formation activity. In the latter cases, the commonly
adopted naming of ``post-starburst'' is in fact misleading.  The exact
duration, time elapsed and strength of the most recent star formation
episode remains hard to constrain, in the absence of high S/N and high
resolution spectra (Leonardi \& Rose 1996). Keeping this in mind, in the
following we will refer interchangebly to ``k+a'' 
or ``post-starburst'' spectra for convenience.

First discovered in distant galaxy clusters over 25 years ago
(Dressler \& Gunn 1983), k+a galaxies have seen a renewed interest in
recent years for a number of reasons.  Being galaxies in rapid
transition from a blue, star-forming phase to a red, passively
evolving phase, they were soon recognized to play an important role in
the evolution of the star-forming fraction in galaxy clusters (the
so-called Butcher-Oemler effect, Butcher \& Oemler 1984), related to
the transformation of spirals in distant clusters into the numerous S0
galaxies that dominate clusters today (Poggianti et al. 1999, Tran et
al.  2003). More recently, it has been realized that the presence of
k+a galaxies in environments other than clusters can contribute to the
evolution of the red and blue galaxy fractions in any environment
(Norton et al. 2001, Tran et al. 2004, Yan et al. 2008)\footnote{At
the time our paper was accepted, the paper from Yan et al. 2008
was available on the astro-ph archive as ``submitted''. Readers should be
aware that, in the following, any comment comparing with Yan et
al. refers to their submitted version, and may not be applicable to
their final, refereed version.}.  At redshifts above 1, k+a spectra
may be a common feature of massive galaxies that formed most of their
stars at high redshift (Doherty et al. 2005, van Dokkum \& Ellis 2003,
van Dokkum \& Stanford 2003).

Post-starburst spectra are observed both in clusters and in the
general field at all redshifts, but their incidence varies with
redshift, environment and galaxy mass.

At intermediate redshifts ($z=0.3-0.6$), several spectroscopic surveys
of galaxy clusters have reported large fractions of k+a spectra
reaching up to 1/4 of the entire galaxy population in distant massive
clusters (Dressler \& Gunn 1983, Couch \& Sharples 1987, Henry \&
Lavery 1987, Fabricant et al. 1991, Dressler \& Gunn 1992, Barger et
al. 1996, Dressler et al. 1999, Tran et al. 2003, 2004, 2007, Swinbank
et al.  2007).  As is obvious, the computed k+a fraction depends
on the adopted criteria for Balmer and [OII] strengths, line
measurement method and spectral quality.  As a consequence, more
significant than the absolute k+a fraction, is the relative fraction
in different environments within the same study.  In general, these
studies have found that the frequency of k+a galaxies is significantly
higher in clusters than in the field at intermediate redshifts
(e.g. Dressler et al. 1999, Tran et al. 2004), although this is not an
universal conclusion: for example, Balogh et al. (1999) argued for a
similarly low k+a proportion in clusters and in the field at $z \sim
0.2-0.3$. Low post-starburst fractions are found both in the field
and in groups at $z \sim 0.8$ from Yan et al. (2008), who argued for 
a 2.5$\sigma$ higher post-starburst fraction in the field than in groups.
A cluster-related origin for a significant population of k+a's at
intermediate redshifts is also supported by the analysis of the
spatial distribution of galaxies with 
k+a spectra in a supercluster region at $z
\sim 0.5$ by Ma et al. (2008), who found that k+a's occur within the
cluster ram-pressure stripping radius and are essentially absent 
in the supercluster filamentary structure.

In the local Universe, k+a spectra are rare among bright galaxies in
{\it all} environments. As a result, since clusters are rare objects,
the majority of k+a galaxies at low-z, in number, reside in the field
(Zabludoff et al. 1996, Blake et al. 2004, Hogg et al. 2006, Goto
2007).  Zabludoff et al. (1996) were the first to note the rather
common association of this type of spectra in the low-z field with
signs of galaxy-galaxy interactions and mergers. Other works have
confirmed these findings, suggesting that the majority of field k+a's
at low-z result from mergers that evolve into early-type galaxies
(Yang et al. 2004, 2008, Goto 2005, Nolan et al. 2007). Their high
merger rate is consistent with the high incidence of spectroscopically
confirmed companions (Yamauchi et al. 2008) and with the fact that
they statistically prefer high local density regions at the scale of
$< 100$ kpc (Goto 2005).

In nearby clusters, ``Balmer-enhanced'' spectra were identified in
several galaxies of the Coma cluster (Caldwell et al. 1993). However,
in the majority of these cases, the Balmer lines were far weaker than
those in distant clusters and Poggianti et al. (2004) concluded that
luminous galaxies with post-starburst/post-starforming spectra are
infrequent in Coma. The latter study revealed a numerous population of
fainter k+a galaxies ($M_B>-17$), comprising $\sim 15$\% of the
cluster dwarfs. If Coma (the only local cluster studied in detail to
date) is representative of massive clusters locally, these results
would imply that the evolution of the k+a luminosity distribution
proceeds in a downsizing fashion (Poggianti et al. 2004).  The same
mechanism may still give rise to cluster k+a's both at high- and
low-z, if the maximum luminosity/mass of actively star-forming
galaxies infalling onto clusters decreases at lower redshift.
A striking correlation between the position of the youngest k+a
galaxies and the substructure in the Coma hot intracluster medium (ICM)
strongly suggests that the quenching of star formation is due to the
interaction of galaxies in infalling groups with the shock fronts in
the ICM.

Several works based on the Sloan Digital Sky Survey (SDSS)
have used the local galaxy
density to investigate the environment of k+a galaxies at low-z
(Balogh et al. 2005, Goto 2005, Hogg et al. 2006, Nolan et al. 2007,
Yan et al. 2008).  They have unanimously found that low-z k+a's do not
preferentially reside in high-density regions, and inferred from this
that they do not preferentially belong to low-z clusters.  However, as
we will show, a low- and intermediate-density preference of k+a's does
not necessarily imply residing outside of clusters. Again from 
SDSS, Yan et al. (2008) found a possibly double peaked density
distribution of post-starburst galaxies, the strongest peak at the lowest
densities sampled and a second peak at densities above average, suggesting
two possible formation channels. Still at low-z, Hogg et al. (2006)
found only a small k+a excess inside the virial radii of clusters, and
a slight excess with near neighbours, supporting the hypothesis that
some k+a spectra are due to interactions with the intracluster medium.

Overall, although a detailed study of a large sample of nearby
clusters is still lacking, the conclusion emerging from all low-z
studies is that there is a paucity of bright, massive k+a galaxies in all
environments, and a suggestive abundance of k+a faint galaxies in
the one studied cluster. In contrast, at intermediate redshifts, most
studies have measured significant fractions of bright k+a galaxies in clusters,
and a higher fraction in clusters than in the field.

It is fundamental to realize that several different physical processes
can produce a k+a spectrum. Any time star formation is truncated on a
short timescale, a galaxy will experience a k+a phase. In different
environments, different mechanisms may be, and very probably are,
responsible for truncating star formation.  The results obtained to
date suggest that mergers may be the predominant mechanism in the
low-z field, while the environmental dependence of k+a's at higher
redshifts and the results in Coma strongly point to cluster-related
phenomena.  Other effects have been proposed, such as AGN and
supernovae quenching (e.g. Bekki et al. 2001, Kaviraj et al. 2007) and
massive dark matter halo formation (Birnboim et al. 2007). In
clusters, besides the aforementioned impact of infalling galaxies with
the ICM shock fronts (Poggianti et al. 2004, Crowl \& Kenney 2006),
other quenching mechanisms include ram pressure stripping (Gunn \&
Gott 1972, Ma et al. 2008, Vollmer 2008), galaxy harassment (Moore et
al. 1996, 1999) and tidal interactions (Koopmann \& Kenney 2004,
Oemler et al. 2008, see also Treu et al. 2003 and Boselli \& Gavazzi
2006 for reviews of cluster processes).\footnote{We note that the
removal of the gas halo reservoir, also known as ``strangulation'' or
``starvation'' (Larson et al. 1980, Balogh et al. 2000, Bekki et
al. 2002) is not expected to produce a k+a spectrum, because star
formation gently declines on a timescale that is comparable or longer
than the A-stars lifetime. As a result, the galaxy does not have a
phase with strong Balmer lines and no emission.}

K+a's are not the only Balmer-strong type of spectra.  Emission-line
spectra with exceptionally strong Balmer lines in absorption (e(a)
spectra) were identified in large numbers at intermediate redshifts,
both in clusters and in the field (Dressler et al. 1999, Poggianti et
al. 1999).
At low-z, this type of spectra are associated with ongoing starbursts
whose young stellar populations are strongly obscured by dust.  In
fact, they are rare among local normal spirals, but common
among dusty starbursts and Luminous Infrared galaxies (Liu \&
Kennicutt 1995, Poggianti \& Wu 2000).\footnote{We note that e(a)
spectra differ from the dusty red sequence galaxies found in the
A901/902 supercluster on the basis of the COMBO-17 photometry (Wolf et
al. 2005, Lane et al. 2007): the latter are mostly optically passive
(low [OII] and low $\rm H\delta$) and have {\it suppressed} star
formation levels and high dust obscuration (Wolf et al. 2007).}  An e(a)
spectrum can be produced if the O and B stars causing the emission lines
are highly obscured by dust and the older A-type stars -- responsible
for the strong Balmer absorption -- have had time to free themselves
or drift away from their dusty cocoons (Poggianti et al. 1999).  On
average, younger stellar generations are more strongly obscured than
older ones, and this phenomenon is often referred to as ``selective
extinction''. As a result of this effect, the strongest starbursts
locally have only weak to moderate emission lines (due to high
extinction), and have usually e(a) spectra, {\it not} spectra
with the strongest emission lines.

Spectrophotometric modeling of e(a) galaxies has confirmed both the
adequacy and the necessity of selective extinction and of an ongoing
starburst to explain simultaneously the optical spectral
characteristics and the FIR fluxes of local e(a) galaxies (Poggianti,
Bressan \& Franceschini 2001, Shioya et al. 2001, Bekki et al. 2001).
Mid-IR ISOCAM (Duc et al. 2002) and Spitzer (Dressler et al. 2008)
studies have confirmed that statistically e(a) galaxies at $z \geq
0.2$ have the highest SFRs, and that they are experiencing an increase
in SFR compared to their past average (Dressler et al. 2008).

From the optical point of view, the only difference between k+a and
e(a) spectra is the presence of emission lines.  If the scenario
outlined above is correct, however, the two spectral classes
correspond to fundamentally different phases of star formation
activity: post-starburst the k+a's, and ongoing dusty starburst the
e(a)'s.  An evolutionary link between the two types of galaxies can
occur, with e(a)'s being in some cases the progenitors of some of the
k+a galaxies (Poggianti et al. 1999, Balogh et al. 2005).

The association between spectral class and star formation history in
these two classes works well in a statistical sense, but the
demarcation between the two histories is not straightforward in all
galaxies and several caveats need to be kept in mind.  
Both radio (Smail et al. 1999) and mid-IR (Dressler et
al. 2008) observations have found that some k+a galaxies may host a
residual ongoing star formation activity, highly obscured by dust and
therefore not visible in the [OII] line.  These works show that k+a
galaxies with ongoing SF are a minority and that the residual SFR is
significantly lower than the SFR during the previous burst (Dressler
et al. 2008).  Moreover, the separation between k+a's and e(a)'s is
commonly set to a low EW of emission lines (usually 5 \AA $\,$ in
[OII]) that is determined by the standard detection limit in distant
spectroscopic surveys. This limit is necessarily arbitrary: lower or
upper detection limits can result in a different mixing between the
two classes. The lines used in the classification and the adopted
limits for their strength vary significantly in the literature and this
obviously hinders the comparison between different works and with the
existing modeling.  Any adopted choice of line combination leads to
some level of bias, incompleteness and possible misclassification
between k+a's and e(a)'s.  For example, the use of the $\rm H\beta$
line in emission has been strongly advocated as a substitute of [OII]
by Yan et al. (2006) to avoid incompleteness in the post-starburst
sample due to AGN-dropouts. In terms of the $\rm
H\beta$-based classification, it is still unclear how
post-starbursts can be separated 
from ongoing dusty starbursts.  Furthermore, e(a)
spectra are sometimes refereed to as ``post-starbursts' in the
literature (eg. Tremonti et al. 2007, see also Le Borgne et al. 2006), 
because the association between
strong Balmer lines and a post-starburst history is common knowledge
in the astronomical community, unlike the association between Balmer
lines and ongoing starbursts. Of course it remains possible that some
e(a)'s are truly post-starburst systems whose lines in emission are
due to a small residual star formation activity (Poggianti et
al. 1999). Finally, spectrophotometric modeling with selective
extinction shows that regular star formation histories can account for
e(a) spectra at $z>0.4$, depending on the timescale of the star
formation decline, without the need to invoke a starburst (Fritz \&
Poggianti in prep.).

Regardless of the intrinsic caveats in the spectral interpretation on
a galaxy-by-galaxy basis, 
the simultaneous
study of emission and Balmer features is the primary source of
information to unveil the recent star formation history of galaxies
observed by spectroscopic surveys: identifying k+a and e(a) spectra allows to
recognize strong variations of the star formation activity (a
truncation or a burst), and to study how they change with $z$ and
environment.  In this paper, we investigate the occurrence and
properties of post-starburst galaxies and dusty starburst candidates
at $z=0.4-0.8$.  We make use of the ESO Distant Cluster Survey dataset
(\S2), adopt the spectral classification described in \S3 and consider
different environments: clusters, groups and the field, as defined in
\S4. We show how the proportion of post-starburst galaxies and dusty
starburst candidates depends on environment (\S5) and how it varies in
clusters with the cluster properties (\S5.2).  We next consider
composite spectra as an alternative tool to study the environmental
dependence of star formation histories (\S6). We present the detailed
properties of galaxies of different spectral types: their morphologies
(\S7.1), masses, luminosities and mass-to-light ratios (\S7.2), local
density and radial distributions (\S7.3).  
Finally, in \S8 we summarize our results and discuss the
implications for the origin and the evolution of post-starburst
galaxies.
In the Appendix, we search for
spectroscopic evidence of galactic winds in our Balmer strong
population.

In the following, we adopt the convention that the EW([OII]3727)
is positive in emission while the EW($\rm H\delta$4101) is positive in
absorption.
All equivalent widths and cluster velocity dispersions are given
in the rest frame. We assume a $\Lambda$CDM cosmology with
($H_0$, ${\Omega}_m$, ${\Omega}_{\lambda}$) = (70,0.3,0.7).

\section{The dataset}

The ESO Distant Cluster Survey (hereafter, EDisCS) is a
multiwavelength survey of galaxies in 20 fields containing galaxy
clusters at $z=0.4-1$.

Candidate clusters were selected as surface brightness peaks in
smoothed images taken with a very wide optical filter ($\sim$
4500-7500 \AA) as part of the Las Campanas Distant Cluster Survey
(LCDCS; Gonzales et al.~2001). The 20 EDisCS fields were chosen among
the 30 highest surface brightness candidates in the LCDCS, after
confirmation of the presence of an apparent cluster and of a possible
red sequence with VLT 20 min exposures in two filters (White et
al. 2005).

For all 20 fields, EDisCS has obtained deep optical photometry with
FORS2/VLT (White et al. 2005), 
near-IR photometry with SOFI/NTT (Arag\'on-Salamanca et al. in prep.), 
multislit spectroscopy
with FORS2/VLT (Halliday et al. 2004, Milvang-Jensen et al. 2008), 
and MPG/ESO 2.2/WFI wide field imaging in $VRI$.
ACS/HST mosaic imaging in $F814W$ of 10 of the highest redshift
clusters has also been acquired (Desai et al. 2007). Other follow-up
programmes include XMM-Newton X-Ray observations (Johnson et
al. 2006), Spitzer IRAC and MIPS imaging (programmes P.I.s Desai,
Finn, Rudnick), $\rm H\alpha$ narrow-band imaging (Finn et al. 2005),
additional FORS2 imaging and spectroscopy in 10 EDisCS fields (Douglas
et al. 2007), 2dF/AAT, VIMOS/VLT and IMACS Magellan wide-field
spectroscopy.

Spectroscopic targets were selected from \textit{I}-band catalogs
(Halliday et al.  2004).  Conservative rejection criteria based on
photometric redshifts (Pell{\'o} et al. 2008) were used in the
selection of spectroscopic targets to reject a significant fraction of
non--members, while retaining a spectroscopic sample of cluster
galaxies equivalent to a purely \textit{I}-band selected one. A
posteriori, we verified that these criteria have excluded at most
1-3\% of cluster galaxies (Halliday et al. 2004 and Milvang-Jensen et
al. 2008).  The spectroscopic selection, observations, and catalogs
were presented in Halliday et al. (2004) and Milvang-Jensen et
al. (2008). Out of the 20 EDisCS fields, we exclude here the two
fields that lack several masks of deep spectroscopy (cl1122.9-1136 and
cl1238.5-1144, Halliday et al. 2004).  For our analysis we use all the
structures in the remaining fields within 0.1 in $z$ from the redshift
targeted for spectroscopy in each field.  The structures used in this
paper are listed in Table~1.

Typically, FORS2/VLT spectra of $>100$ galaxies per cluster field were
obtained, with exposure times of 4 hours for the high-z sample and 2
hrs for the mid-z sample. The analysis presented in this paper is
based on these data. 

Given the long exposure times, the success rate of our spectroscopy
(number of redshift/number of spectra taken) is 97\% above the
magnitude limit used in this study (see Poggianti et al. 2006 for
details).  The completeness of our spectroscopic catalogs, which depends on
galaxy magnitude and distance from the cluster center, was 
computed for each cluster in Poggianti et al. (2006). This is used
in this paper to weight the contribution of each galaxy and therefore
correct the sample for spectroscopic incompleteness. Typically,
our spectroscopy samples a region out to a clustercentric radius
equal to $R_{200}$. This is defined to be the radius delimiting a sphere
with interior mean density 200 times the critical density and is
commonly used as an approximation for the cluster virial radius. The
$R_{200}$ values for our structures are computed from the velocity
dispersions using eqn.~1 in Poggianti et al. (2006).

The EDisCS spectra have a dispersion of 1.32 \AA $\,$ pixel$^{-1}$ or
1.66 \AA $\,$ pixel$^{-1}$, depending on the observing run.  They have
a FWHM resolution of $\sim 6$ \AA, corresponding to rest-frame 3.3 \AA
$\,$ at z=0.8 and 4.3 \AA $\,$ at z=0.4.  The equivalent widths of
\oii and $\rm H\delta$ were measured from the spectra using a gaussian
line-fitting technique, as outlined in Poggianti et al. (2006), using the
same technique of Dressler et al. (1999).  Our measurements
are based on the visual inspection of each 1D and 2D spectrum.  Each
emission line detected in a given 1D spectrum was confirmed by visual
inspection of the corresponding 2D spectrum. This method is especially
useful for assessing the reality of weak \oii lines, to avoid
false positive and false negative detections of [OII] emitters. 
Errorbars on EWs were estimated allowing continuum variations according
to error spectra computed as in Sanchez-Blazquez et al. (2008).
\footnote{Given the interactive method of classification, a Monte
Carlo approach for the treatment of errors, recommended in the case of
purely automated line measurements, would be inappropriate in our
case.} 

We do not attempt to separate a possible AGN contribution to the [OII] line,
or to exclude galaxies hosting an AGN.  We are unable to identify AGNs
in our data, since the traditional optical diagnostics are based on
emission lines that are not included in the spectral range covered by
most of our spectra (see Poggianti et al. 2008 and Sanchez-Blazquez et 
al. 2008 for a discussion of AGN occurrence in our red sample).

Only galaxies with an absolute $V$ magnitude brighter than $M_V^{lim}$
are considered in the following. $M_V^{lim}$ varies with redshift
between -20.5 at $z=0.8$ and -20.1 at $z=0.4$ to account for passive 
evolution. In the following we will refer to this as the magnitude-limited
sample.

\section{The spectral types and their interpretation}

In the following, we adopt a spectral classification similar to that
proposed by the MORPHS collaboration (Dressler et al. 1999). This
method of classification is based on the equivalent widths of the
[OII] line in emission and the $\rm H\delta$ line in absorption, which
are the best indicators of ongoing and recent star formation,
respectively, in optical spectra of galaxies at $z=0.4-1$.

The classification allows us to broadly divide galaxies into groups according to
their star formation history. In particular, it aims to identify
galaxies with an ongoing (starburst) and a recent (post-starburst)
episode of star formation, to distinguish them from passively
evolving galaxies and quiescently star-forming galaxies.  This scheme
is based upon and is a development of earlier spectral interpretation of strong
Balmer absorption in the spectra of galaxies in intermediate redshift
clusters (Dressler \& Gunn 1983, Couch \& Sharples 1987).

The classes that we will use in the following are:

a) {\it Passively evolving galaxies - k class} showing no spectral
sign of ongoing or recent star formation during the past 1-1.5 Gyr.
We define them as those with no securely detected [OII] line in
emission and EW$(\rm H\delta)<3$ \AA $\,$ and refer to them as ``k''
galaxies (with a spectrum similar to a K-type star).  In the
following, we will refer occasionally to ``k(e)'' galaxies, which
correspond to those with an otherwise k-type spectrum and an
additional very weak [OII] line in emission ($<5$ \AA), whose reality
is confirmed by the 2D spectrum.

b) {\it Post-starforming/post-starburst galaxies - k+a class}.  
Spectrophotometric modeling has shown that the
lack of significant [OII] emission (EW$<5$ \AA) and the strength of
$\rm H\delta$ (EW$>3$ \AA) imply that the star formation activity
was terminated sometime between $5 \times 10^7$ and $1.5 \times 10^9$ years
prior to observation.  We refer to these as ``k+a'' (a mix of K-type
and A-type stars) galaxies.

c) {\it Quiescent star-forming galaxies - e(c) class}, which have
spectra with a moderate [OII] line in emission (EW=5-25 \AA) and an
EW$(\rm H\delta)<4$ \AA.  These spectral characteristics are similar
to those of normal spiral galaxies of types Sa to Sd in the local
Universe. They are consistent with a continuous star formation history
during which a galaxy experienced no sudden variation (either
starburst or SF-truncation) in its star formation activity.

d) {\it Dusty starburst candidates - e(a) and e(a)+ classes.} E(a)'s
are defined to have an EW(OII)$>5$ \AA $\,$ and a strong $\rm H\delta$
line in absorption (EW$(\rm H\delta)>4 $ \AA). As discussed in \S1, at
low redshift these line strengths are observed to be rare among normal
spirals, while they are common among dusty starbursts and Luminous
Infrared galaxies. These observations and spectrophotometric modeling
of e(a) galaxies suggest that this spectral combination arises in
dusty, ongoing starbursts. As discussed in \S1, in principle
the e(a) signature can arise from truly post-starburst systems
with a residual star formation activity, although examples of such a
case have not yet been securely identified observationally in galaxy 
clusters.

$\rm H\delta$-strong galaxies such as e(a)'s typically also have 
strong higher-order Balmer lines, i.e. strong $\rm H\epsilon$
(epsilon, 3970 \AA), $\rm H\zeta$ (zeta, 3889 \AA), $\rm H\eta$ (eta,
3835 \AA) and $\rm H\theta$ (theta, 3798 \AA), as shown in the bottom
right panel of Fig.~\ref{spectra}.\footnote{$\rm H\eta$ (eta) can be
very strong also in old, metal-rich stellar populations, e.g. see
Dressler et al. 2004 and our k composite spectrum in Fig.~\ref{spectra}.} 
The line that correlates most strongly with $\rm H\delta$ and is the best indicator
of recent star formation among the high-order Balmer lines is $H\zeta$
(zeta), that is a useful substitute for $\rm H\delta$ 
whenever the latter is unavailable.

In EDisCS spectra, we noted a large number of galaxies that have
EW$(\rm H\delta)<4$ \AA $\,$ and therefore strictly speaking are not
classifiable as e(a)'s, but have very strong higher-order Balmer lines
similar to those of e(a)'s. An operational definition to categorize this type
of spectra 
is to select those with $H\zeta>4$ \AA.  In the following we will
refer to these as ``e(a)+'' spectra.  Their similarity to e(a)'s is
visible comparing the two bottom panels in Fig.~\ref{spectra}.

E(a)+ spectra only differ from e(a)'s because the $\rm H\delta$ line
in absorption is slightly more filled in by emission.  In the
following we consider them separately from both e(a)'s and e(c)'s, to
allow for the possibility that at least some e(a)+ galaxies may be
candidate dusty starbursts too.

e) {\it Strong emission line starbursts - e(b) class}, defined to have
an EW$(OII)>25$ \AA. Both in the local Universe and in clusters at
$z\sim 0.5$, galaxies with a strong [OII] line are generally low-mass,
very late-type galaxies (Poggianti et al. 1999). In \S7 we show
that this is the case also in the EDisCS sample. These spectra belong to
``starbursts'' in the sense that their current SFR is much higher than
their past average, but they can be either starbursts or simply
galaxies with a regular star formation rate of a late-type
and low extinction levels.

Following Dressler et al. (1999), we present in Fig.\ref{spetypes}
a summary view of the equivalent width limits adopted to define
the spectral types described above.

\subsection{Robustness of the spectral type interpretation}

\begin{figure*}
\vspace{-0.1cm}
\centerline{\hspace{1cm}\includegraphics[width=1.0\columnwidth]{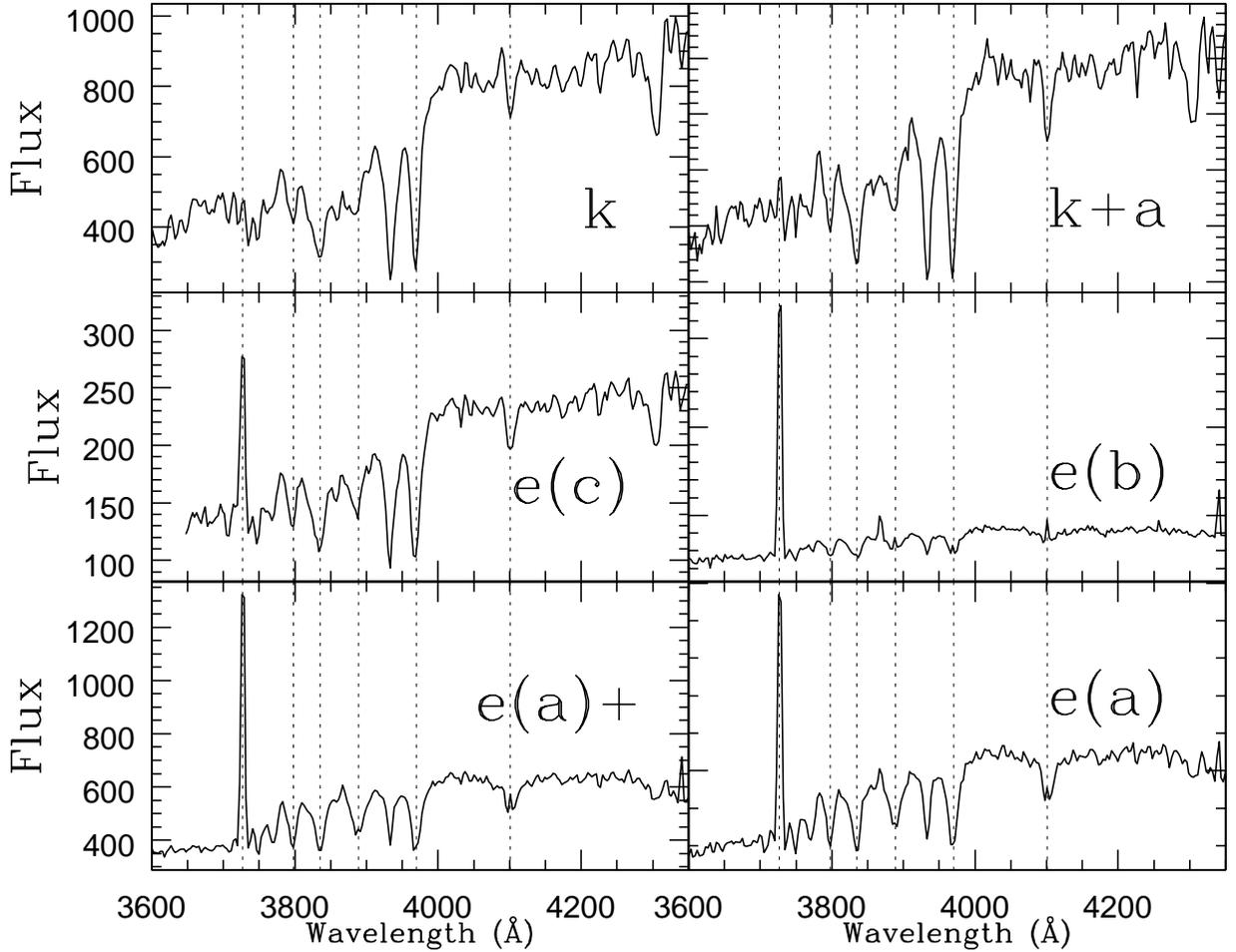}\hfill}
\caption{Composite spectra of the main spectral classes described in
\S3.  Spectra are given in the rest frame, rebinned at 3 \AA $\,$ for
displaying purposes. Only a limited spectral region including [OII]
and $\rm H\delta$ is shown. The main lines of interest are: [OII]3727,
$\rm H\delta$4101, $\rm H\epsilon$3970 (epsilon), $\rm H\zeta$3889
(zeta), $\rm H\eta$3835 (eta) and $\rm H\theta$3798 (theta).
 \label{spectra}}
 \end{figure*}

\begin{figure*}
\vspace{-0.1cm}
\centerline{\hspace{1cm}\includegraphics[width=1.0\columnwidth]{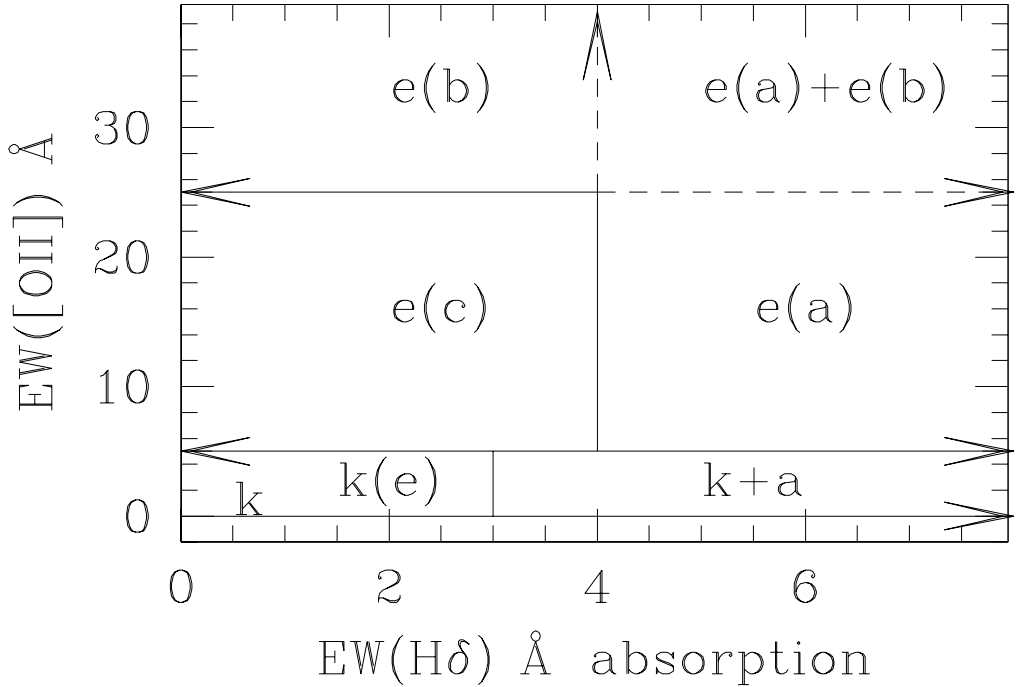}\hfill}
\caption{Schematic representation of the spectral classification scheme 
described in \S3. The diagram shows for each spectral type the range in 
rest frame [OII] emission and $\rm H\delta$ absorption equivalent widths.
Note that the ``k'' type is defined to have no securely detected [OII] 
emission, therefore corresponds to EW([OII])=0.
 \label{spetypes}}
 \end{figure*}

To visualize the variation in spectral properties between the classes
described above, we show the composite spectrum of the main spectral
classes in Fig.~\ref{spectra}, zooming in on the spectral region
containing [OII] and $\rm H\delta$. Each composite spectrum is
obtained by coadding the individual galaxy spectra of that class, after
normalizing each spectrum by its mode and assigning a weight equal to
its observed I-band luminosity.  The EWs of [OII] and $\rm H\delta$
measured on the composite spectra are given in Table~2.

The differences between the Balmer and [OII] properties in the
composite spectra of different spectral types are visible in
Fig.~\ref{spectra}, thanks to the high signal-to-noise of the
coadded spectra. Single galaxy spectra are of course noisier, and
therefore more prone to classification errors.  The visual inspection
of each spectrum alleviates significantly this problem, in two ways:
a) the comparison between the 1D and 2D spectrum provides a 
reliable detection/non-detection of even weak emission and b) the visual
assessment of the quality of the Balmer measurement in most cases is able to
distinguish noisy or problematic (due e.g. to sky lines)
measurements of $\rm H\delta$ from reliable line detections and
measurements.  Our classification is therefore based both on the EW
limits given in \S3, and on the visual assessment of the quality of
the $\rm H\delta$ strength and its confirmation by the strength of the
higher order Balmer lines.
When a spectral type could not be reliably determined, a
``?''-type was assigned (Table~3).  Similarly, if the Balmer strength
could not be reliably measured in an emission-line spectrum, an
``e''-class was assigned (Table~3).

The robustness of the spectral classification is 
confirmed by the fact that repeated observations with a securely
assigned spectral type always yielded the same spectral
classification.  Moreover, we performed a test on the differences
between the two spectral classes that in principle are most prone to
misclassification between each other, i.e. the k and k+a spectra.  We
created a k+a ``bootstrap sampled'' composite by coadding an identical
number of spectra as for the original composite in Fig.~\ref{spectra}, 
but using a
random selection of k+a spectra from the original list, allowing a
spectrum to be selected more than once.
All of the 100 bootstrapped composites of
k+a galaxies have an $\rm H\delta$ (and $\rm H\zeta$) strength that is
incompatible with (significantly higher than) the k-composite strength.

The reliability of the k+a classification is particularly important in
order to avoid a ``spillage'' from the k into the k+a class due to
noisy measurements. The spectral visual assessment, coupled with the
quality of the spectra, is especially useful to eliminate
spurious cases and select a reliable k+a sample. As we will show later
in the paper, the reality of the differences between our k and the k+a
galaxies is testified, a posteriori, by their different morphological
and clustercentric radial distributions. Furthermore, we will show
that cluster composite spectra, that are free from any spillage
effect, lead to the same conclusions as the analysis of the frequency
of k+a spectra in each cluster.

Yan et al. (2006) argued that, locally, about half of the red galaxies
with emission lines have line ratios typical of LINERs. Assuming
LINERs do not host any ongoing star formation activity, this would
imply that about half of the red, emission-line galaxies would be
erroneously classified as ``star-forming'' due to the presence of
AGN-powered emission. As a consequence, a post-starburst galaxy with
some AGN-powered emission could fail to be assigned to the
post-starburst sample, and the latter would result incomplete.

In our sample, all red, emission line galaxies have {\it weak}
Balmer ($\rm H\delta$ and higher order) lines, hence would not be
classified as k+a's even in absence of emission.  The only exception
are 2 red emission-line galaxies with strong Balmer absorption, which
correspond to 0.3\% of our sample.  Incompleteness in our k+a sample
due to pure AGN contamination is therefore not an issue.

\section{Defining environments}
The EDisCS dataset allows us to study galaxies in a wide range of environments
using homogeneous data. In this paper we consider:

1) {\it Clusters}, defined as structures with a velocity dispersion
$\sigma > 400 \, \rm km \, s^{-1}$. Our clusters, listed in Table~1,
span the entire range of cluster masses with velocity dispersion
from $400$ to $1100 \, \rm km \, s^{-1}$.

2) {\it Groups} are structures with $160 < \sigma < 400 \, \rm km \,
s^{-1}$ with at least 8 spectroscopic members.  Groups are further
split into groups with high- and low-OII content, depending on their
fraction of [OII]-emitters. {\it High/Low-OII groups have an [OII]
fraction higher than 80\% and lower than 50\%, respectively.}  

The high-OII groups and the clusters define a broad anticorrelation in
a diagram of [OII] fraction versus cluster velocity dispersion
$\sigma$ ($f_{[OII]} = -0.74 (\sigma /1000) + 1.115)$). Low-OII groups
are ``outliers'' in this relation (Fig.~4 in Poggianti et al. 2006).
In the following, we will refer collectively to "[OII]-outliers" as
those groups and clusters that have a low [OII] content for their
$\sigma$ compared to the rest of the structures.  Specifically, the
outliers have an [OII] fraction lower by at least 25\% than the fraction
derived from the best fit relation at its velocity dispersion, thus
$f_{[OII]}< -0.25 + (-0.74 (\sigma/1000) + 1.115)$ (see Fig.~10 and
eqn.2 in Poggianti et al. 2006).

3) {\it Poor groups} are galaxy associations with between 3 and 6 galaxies,
identified as described in \pog. No velocity dispersion
measurement is attempted for them.

4) The {\it ``field''} includes those galaxies that do not belong to
any cluster, group or poor group.

Only groups, poor groups and field galaxies within $\pm0.1$ in
redshift from the cluster targeted in each field were considered. With
this choice, the spectroscopic catalog of each environment is
equivalent to a purely I-band selected sample with no selection bias,
although it should be kept in mind that there may be a preference
for our field galaxies to lie in the filaments close to our clusters and 
groups.

\section{Results: the star formation histories in different environments}

\begin{figure*}
\centerline{\hspace{3cm}\includegraphics[width=1.2\columnwidth]{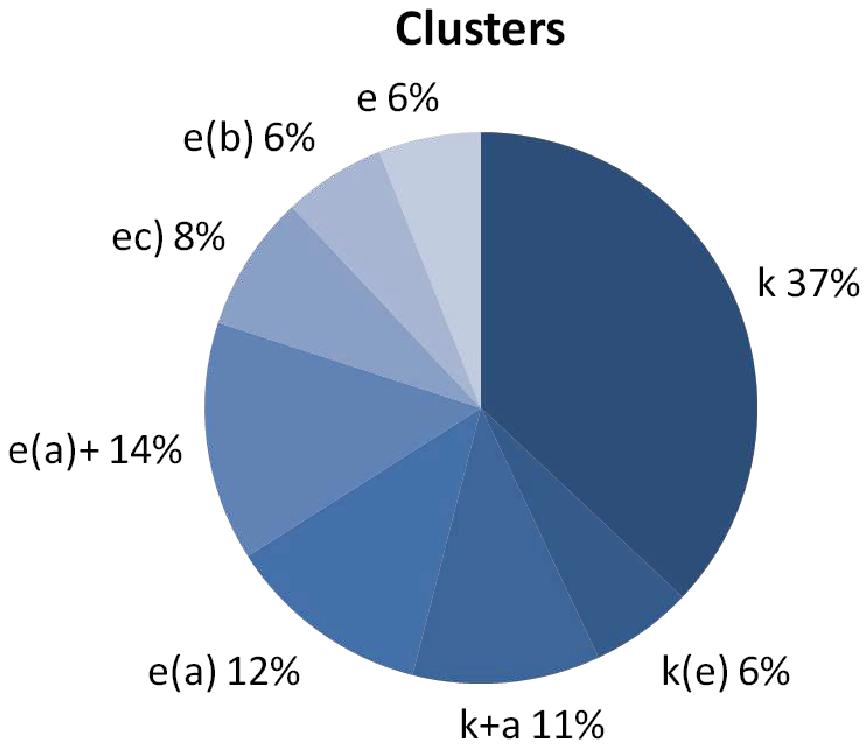}\hfill\hspace{-12cm}\includegraphics[width=1.2\columnwidth]{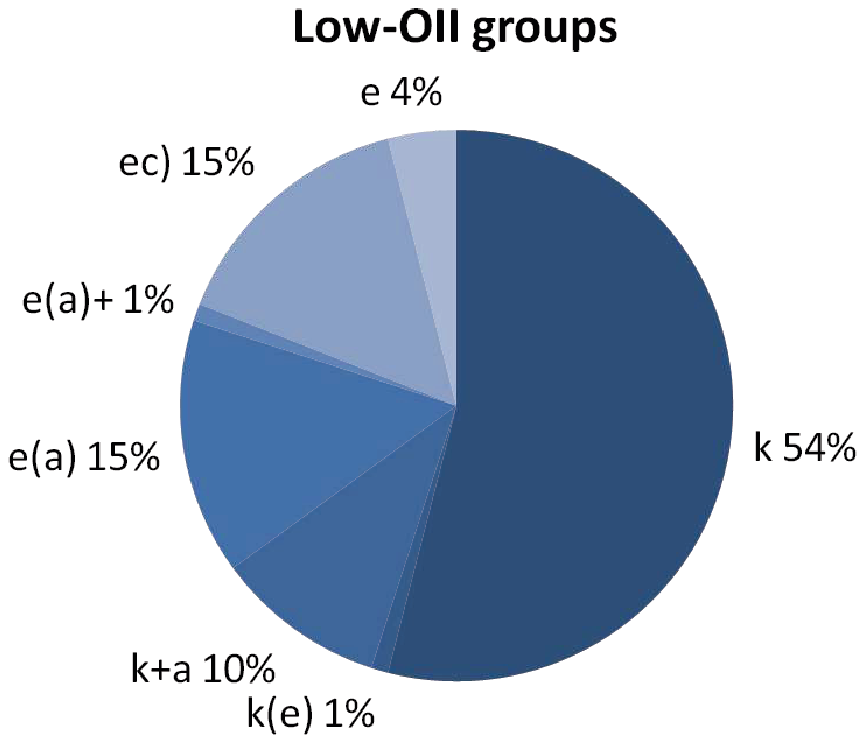}}
\vspace{-9cm}
\centerline{\hspace{3cm}\includegraphics[width=1.2\columnwidth]{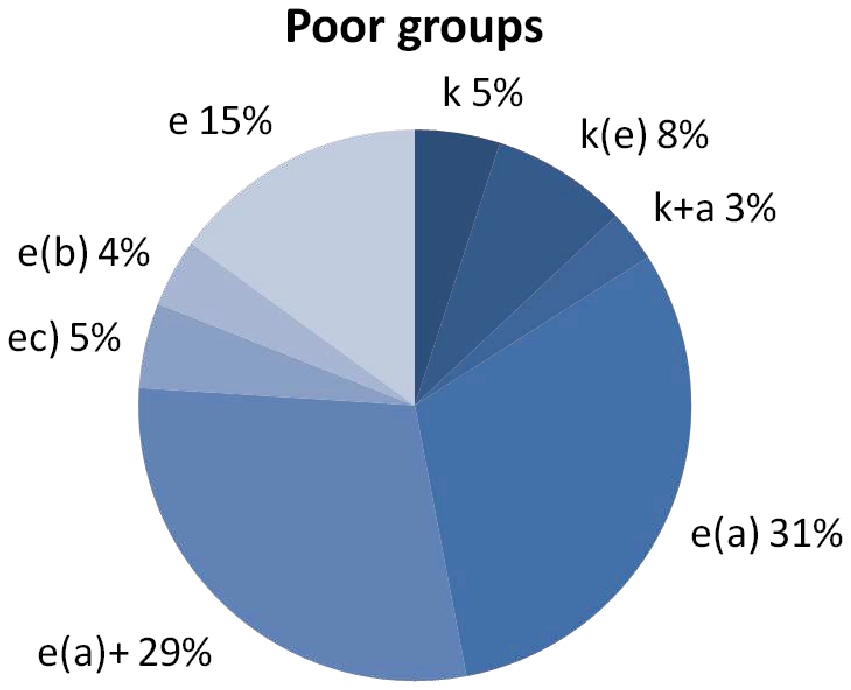}\hfill\hspace{-12cm}\includegraphics[width=1.2\columnwidth]{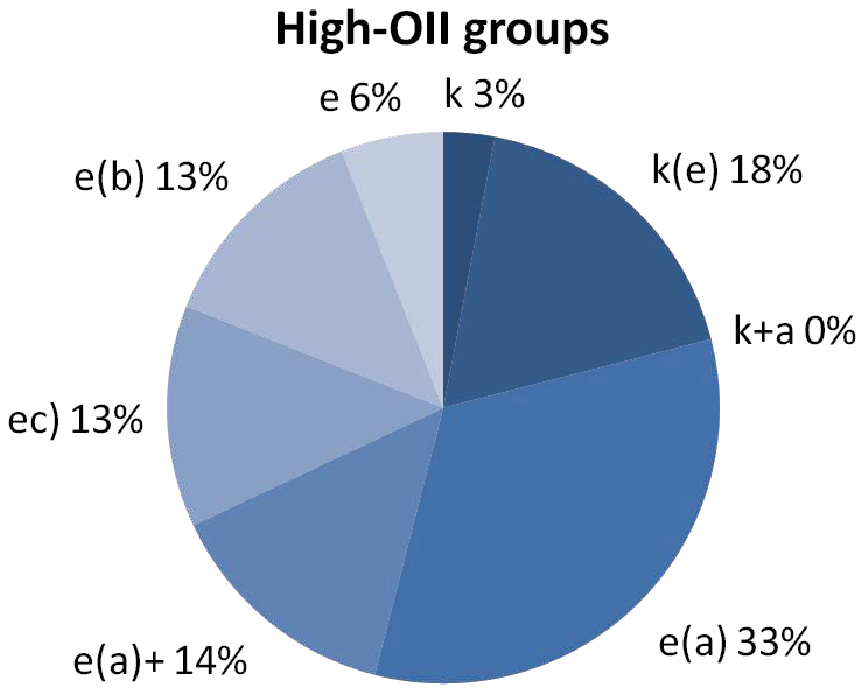}}
\vspace{-9cm}
\centerline{\hspace{3cm}\includegraphics[width=1.2\columnwidth]{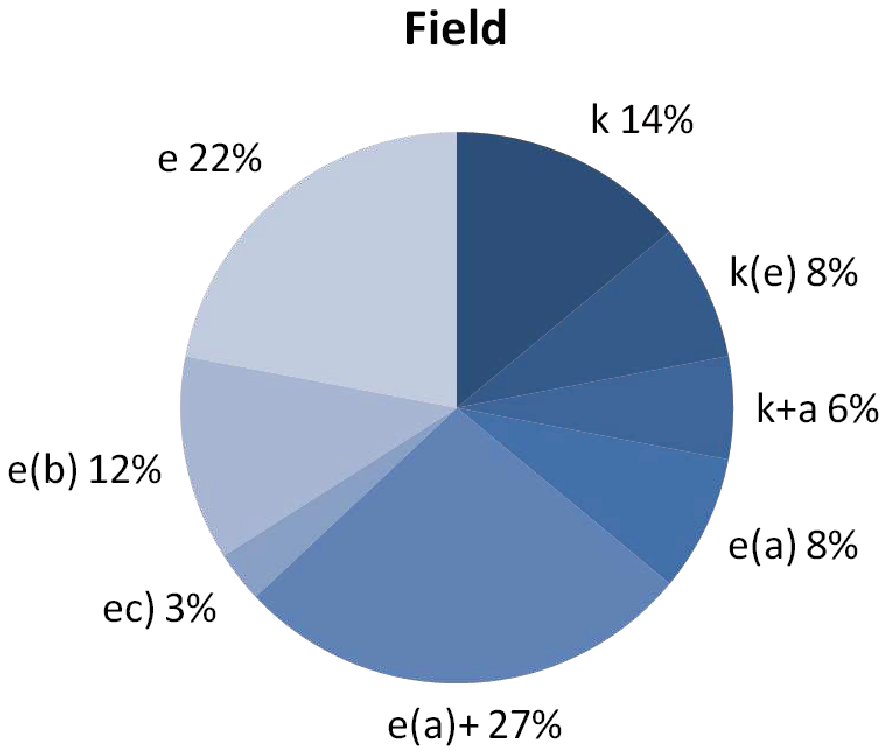}\hfill}
\vspace{-9cm}
\caption{Graphical representation of the spectral fractions in
different environments computed within $R_{200}$ and corrected for
spectroscopic incompleteness, as in Table~4.
 \label{pi}}
 \end{figure*}

We begin our analysis by examining the relative fractions of each
spectral type in the different environments. The raw fractions,
measured for the whole magnitude-limited spectroscopic sample with no
radial constraints, are given in Table~3. Table~4 lists the fractions
for the magnitude-limited sample weighted for spectroscopic
incompleteness and considering only galaxies within $R_{200}$.  The
results of Table~4 are visually illustrated in Fig.~\ref{pi}.

Tables~3 and ~4 refer to a magnitude-limited sample of galaxies.  We
also define a mass-limited sample including, in all environments, only
galaxies with stellar masses $>5.0\times 10^{10} \, M_{\odot}$. This
is the stellar mass of a galaxy with luminosity $M_V=-20.3$ (our
magnitude limit at the average redshift) that has the highest
mass-to-light ratio we observe in our sample, $M/L_V = 4 \,
M_{\odot}/L_{\odot}$ (\S7.2).  Once corrected for spectroscopic
incompleteness, our sample is complete above this mass limit for all
spectral types and all redshifts.  Galaxy stellar masses and absolute
luminosities were computed as described in Poggianti et
al. (2008).  The fractions of each spectral type for the mass-limited
sample are presented in Table~5.

As far as the post-starburst galaxies are concerned, the tables 
and Fig.~\ref{pi} show
that the proportion of k+a galaxies is comparably high in clusters
and in low-[OII] groups (about 10\%), while 
the poor groups and the field contain
fewer post-starburst galaxies, and no k+a galaxy
is detected in the groups with high [OII]. 

This has two main implications.  First of all, these results
highlight, once more, the differences in the galaxy populations of
groups of similar velocity dispersion, but different [OII] content. We
have noted before that the properties of galaxies in low-[OII] groups
resemble those in the cores of far more massive clusters, in terms of their
fractions of passively evolving and early-type galaxies, their [OII]
EW distribution (Poggianti et al. 2006) and their steep relation
between average star-forming fraction and local galaxy density
(Poggianti et al. 2008). We now find that these systems are 
similar to more massive clusters also for the conspicuousness of their
post-starburst population. In contrast, high-[OII] groups of similar
masses appear to be the least-favourable environment for the production of
post-starburst galaxies.

Second, post-starburst galaxies are observed for the first time to
occur in significant proportions also in a subset of the groups, while
so far, at $z>0.4$, they were believed to be a phenomenon related
principally to massive clusters. As discussed in \S1, previous work
found a low k+a fraction in the general field and a high k+a fraction
in clusters at $z=0.4-0.5$. Intermediate and low-mass clusters and
groups at these redshifts had not been studied in detail to date.
Hints of a sizeable post-starburst population in low-z groups were
found by Zabludoff \& Mulchaey (1998). Our result shows that a
mechanism switching off star formation on a short timescale must
operate efficiently also in {\it some} (but not all) groups.

We note that the k+a fraction computed over all groups (low- and
high-[OII] groups together) is low, of the
order of that computed in the field (see line ``All groups'' in the
tables).  This might explain the recent result of Yan et al. (2008),
reporting low k+a fractions in the DEEP2 field and groups at $z=0.8$ and a
2.5$\sigma$ higher fraction in the field than in groups. Given the
different environment and post-starburst definitions, a straight
comparison cannot be performed, but their group sample is likely
dominated by a combination of our poor groups, high-[OII] groups and
low-[OII] groups: our poor group and overall group k+a fractions
(e.g. 0.03 and 0.05, respectively, see Table~4) and field fractions
(0.06) are probably qualitatively consistent with their result.

It is also of note that 
the k+a fraction we observe in clusters is significantly lower than
that measured by the MORPHS collaboration, the only previous large
spectroscopic survey of several distant clusters at redshifts similar
to ours (Dressler et al. 1999). The MORPHS work has the same EW
measurement method and spectral classification, and similar spectral
quality, galaxy magnitude and radial limits of the present work,
therefore a direct comparison is appropriate.  In 10 clusters at
$z=0.4-0.5$, the MORPHS reported a total k+a fraction of $\sim 20$\%
(Poggianti et al. 1999), which is approximately twice the value we observe
in EDisCS.  In the following, the analysis of the detailed cluster
properties will shed light on the origin of the different k+a
incidence in the two samples (\S5.2).

The conclusions discussed above are valid both for the
magnitude-limited and mass-limited samples. Figure~\ref{pi} and
tables~3, ~4 and ~5 also show that the environments with the highest
k+a fraction (i.e. clusters and low-[OII] groups) are those with the
highest fraction of passively evolving (k) galaxies, and lowest
fractions of actively star-forming (emission-line) galaxies.  At
$z=0.4-0.8$, those environments that host few currently star-forming
galaxies are also efficiently truncating star formation in them.

\subsection{Quenching efficiency and starburst frequency}

Each galaxy in our sample is observed in one of three main
evolutionary phases: while it is passively evolving (k/k(e)), recently
star-forming (k+a) or currently star-forming (any of the emission-line
classes).  It is useful to examine the incidence of post-starburst
galaxies within the ``active'' population, defined to include k+a
galaxies and galaxies of all emission-line types. The active
population represents the population of galaxies that were all
star-forming $\sim 2$ Gyr before the epoch of observation.

Hence, the k+a/active fraction can be considered a sort of
``quenching efficiency'': the efficiency of a given
environment in truncating star formation in star-forming galaxies,
which have presumably been recently accreted by the cluster.\footnote{Given   
the evolution with redshift of the star-forming fraction
in clusters (Poggianti et al. 2006, and all references concerning the
Butcher-Oemler effect), it is reasonable to assume that sooner or later a
star-forming cluster galaxy turns into a passively evolving one, and to
identify active galaxies with the population of recently star-forming
{\it and} recently accreted galaxies, that are observed before they
turn passive.}

\begin{figure*}
\vspace{-0.1cm}
\centerline{\hspace{1cm}\includegraphics[width=1.0\columnwidth]{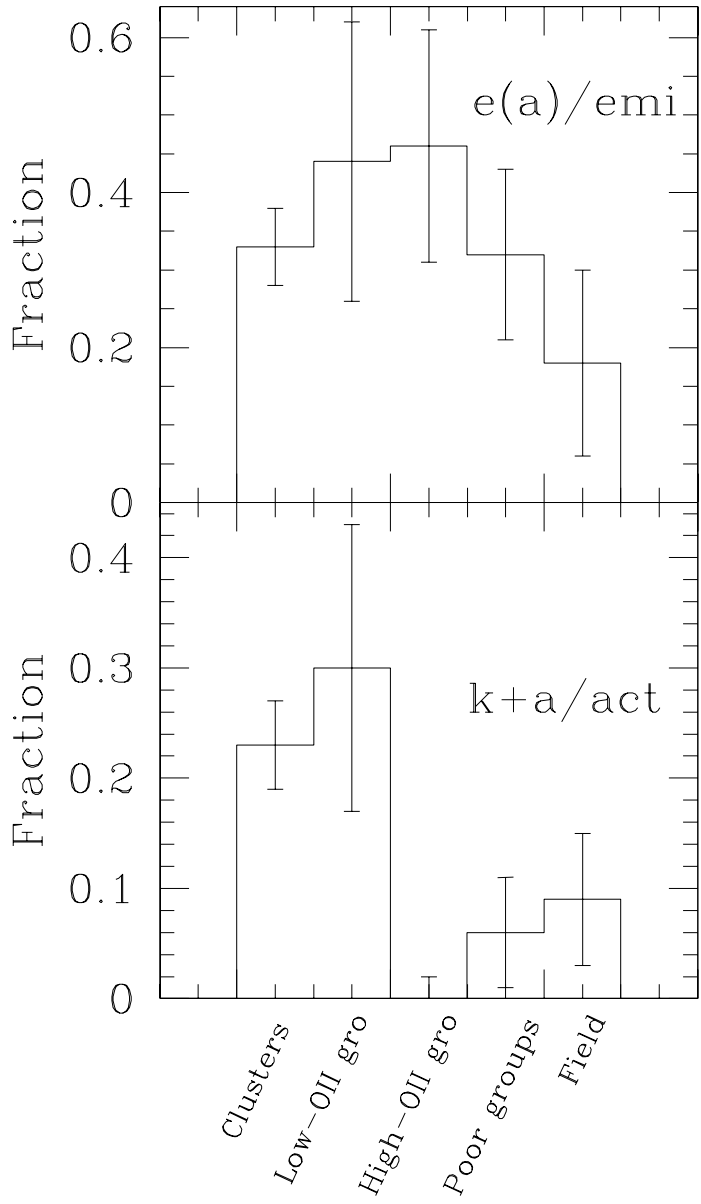}\hfill}
\caption{K+a/active fractions and e(a)/emission fractions in the different
environments, as in Table~6.
 \label{histo}}
 \end{figure*}

The environmental variations of the star formation histories are even
more striking when considering the spectral fractions relative to the
active population instead of the fractions in the entire cluster
population.  The k+a/active fractions are listed in Table~6 and shown
in Fig.~\ref{histo} for the
various EDisCS environments.  Post-starburst galaxies represent 20\%
to 30\% of the active galaxy populations in EDisCS clusters and
low-[OII] groups, while they compose only less than 10\% of the active
galaxies in the other environments.

The same table shows also a comparison with the MORPHS cluster and
field results at $z=0.4-0.5$. The k+a/active incidence is even higher
in MORPHS than in EDisCS clusters (40\% versus 23\%), and it is equal
in the EDisCS and MORPHS field (9\%).

Considering now the dusty starburst candidates, Table~6 lists both the
e(a)/active fraction and the fraction of e(a)'s relative to the number
of emission-line galaxies with an assigned spectral type (the sum of
e(a)'s, e(a)+'s, e(c)'s and e(b)'s).  The e(a)/emission fraction,
graphically displayed in Fig.~\ref{histo}, is related to the
occurrence of dusty starburst candidates among star-forming galaxies,
whose star-formation has not yet terminated.

Previous works (Poggianti et al. 1999, Dressler et al. 1999, 2004,
2008) observed a high frequency of e(a) galaxies both in clusters
and in the field at $z=0.4-0.5$, suggesting that the presence of a
dusty starburst is not caused by the cluster environment (see also Ma
et al. 2008).  Table~6 shows that the e(a)/active and e(a)/emission
fractions observed in the EDisCS and MORPHS clusters and field are
very similar.

Besides confirming the ubiquity of e(a) spectra in all environments at
$z=0.4-0.8$, we find that the highest e(a) frequency is observed in
groups ($\sim 45$\% of the star-forming galaxies), while lower
fractions (20\% to 30\% of the star-forming galaxies) are observed in
clusters, poor groups and the field.  Interestingly, the e(a)/emission
fraction is high in {\it all types} of groups, being similarly high in
low- and high-[OII] groups.
Groups with $\sigma =150-400 \, \rm km \, s^{-1}$ are therefore the
most favourable environment for triggering an e(a) spectrum.  The
frequency of dusty starburst candidates in groups could be related
to the high incidence of mergers expected in these environments.

The results presented so far highlight the strong environmental
dependence of the frequency of k+a spectra, and a milder but
noticeable variation in the proportion of dusty starburst candidates
with environment. It is worth noting that even in a mass-limited
sample the different mix of star formation histories in the various
environments could arise from a different distribution of galaxy
masses with environment (see Poggianti et al. 2008). In order to test
whether our results are driven by this effect, we constructed a
mass-matched sample, drawing, for each cluster galaxy, a galaxy of
similar mass in each of the other environments. All conclusions
presented in this section remain unchanged: strong environmental
variations in the spectral fractions persist in the mass-matched
samples. Hence, variations in the galaxy mass distribution are not
responsible for the observed variations in spectral fractions with
environment.  Galaxies of similar masses must present significantly
different star formation histories depending on environment.

\subsection{Dependence on cluster properties}

In \S5.0 and 5.1 we have shown that the fractions of
post-starburst and, to a smaller extent, of dusty starburst candidates
depend on the global environment, when environment is 
subdivided into clusters, different types of groups, poor groups
and the field. 
To investigate whether the star formation histories depend on the
cluster and group properties in more detail, we now examine how
the spectral fractions depend on the velocity dispersion of the
system. The latter can be considered to be a proxy for the system mass,
being $M \propto {\sigma}^3$ at a fixed redshift.

\begin{figure*}
\centerline{\hspace{1cm}\includegraphics[width=0.5\columnwidth]{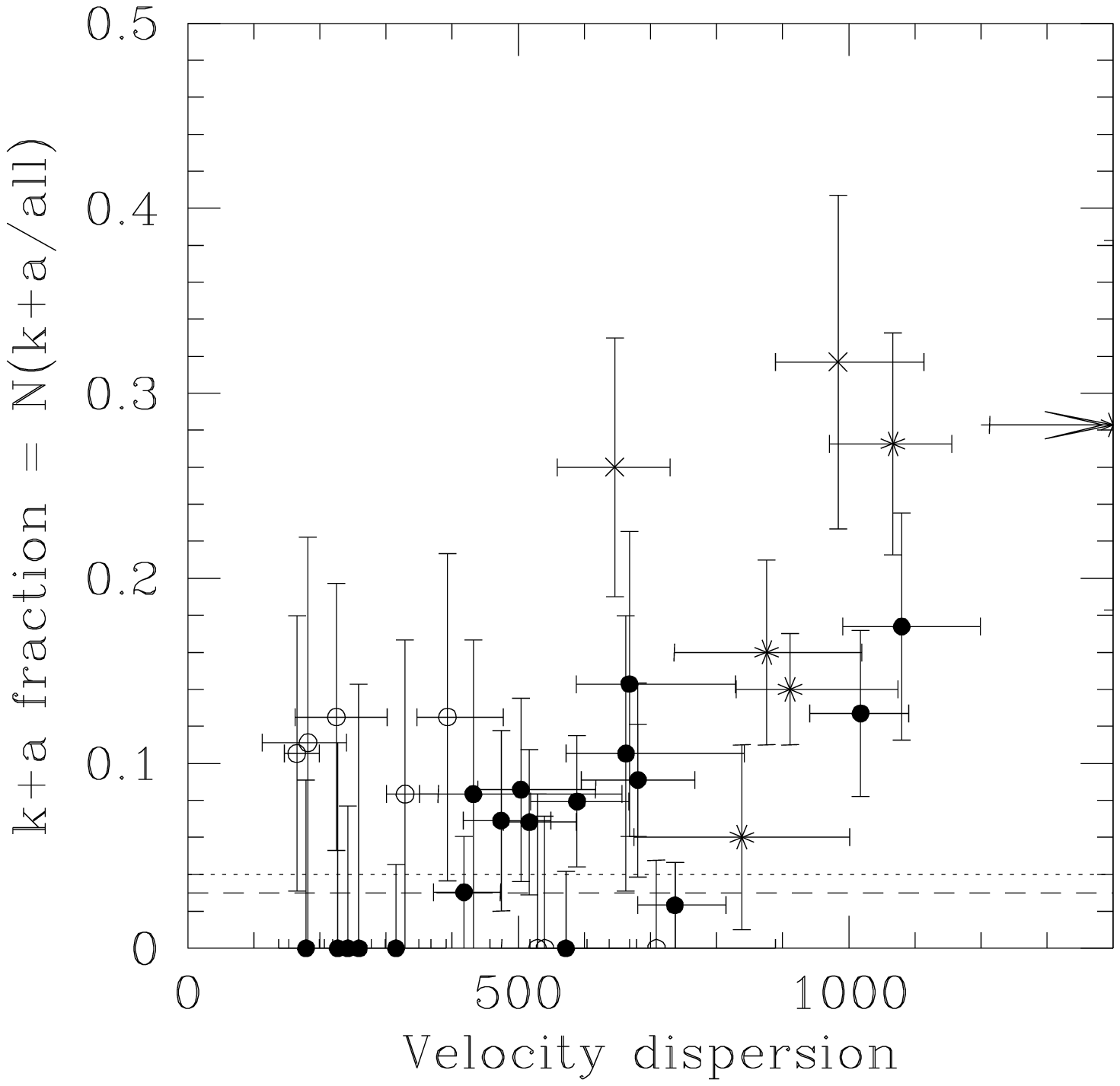}\hfill\hspace{-2cm}\includegraphics[width=0.5\columnwidth]{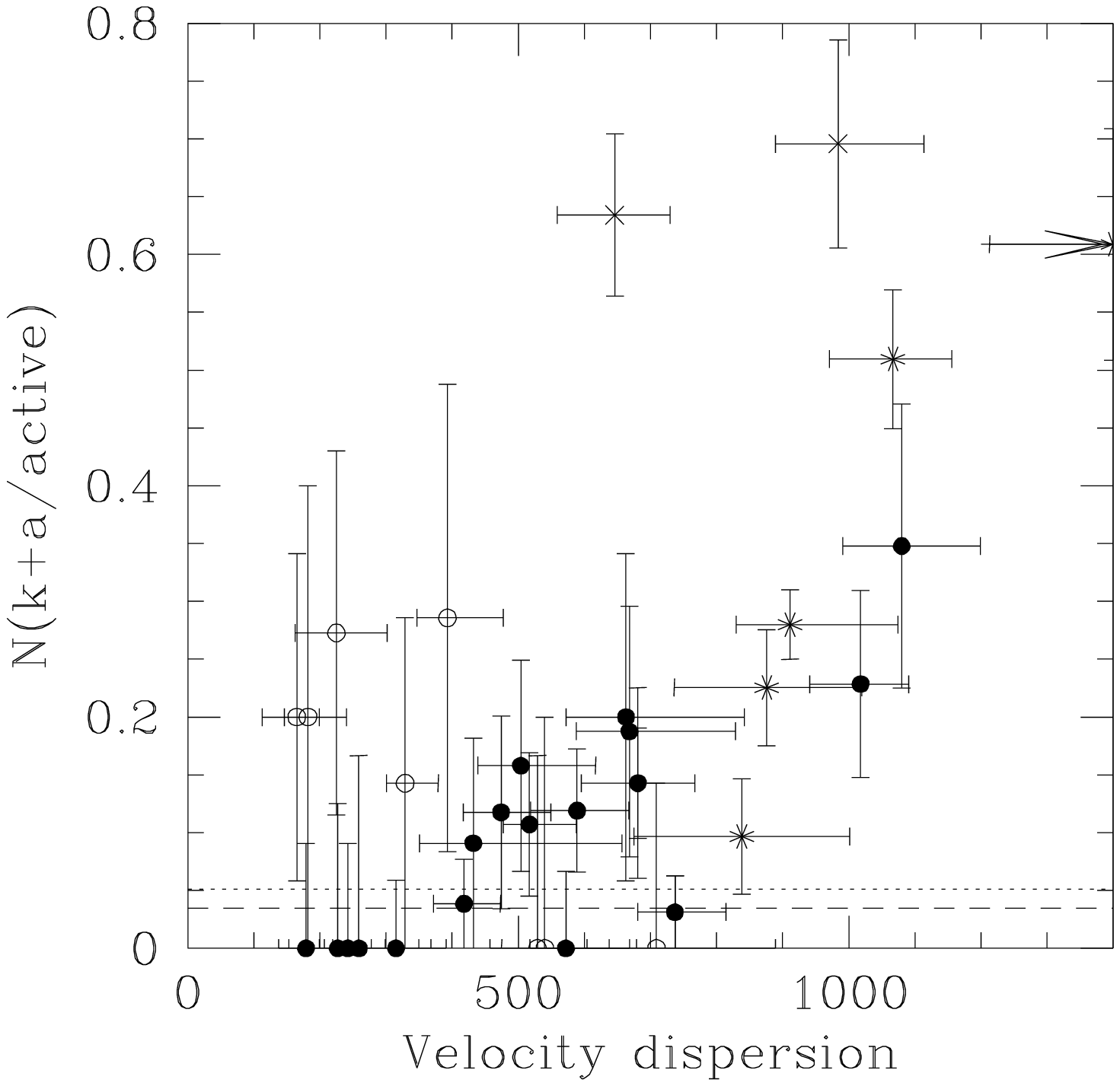}}
\caption{Fraction of k+a galaxies among the whole cluster population
(left) and among the active population (right) versus cluster velocity
dispersion. The k+a/active fraction shown in the right panel
represents the ``quenching efficiency'' (see \S5.1).  Circles are
EDisCS structures, outliers in the [OII]-$\sigma$ relation are indicated
by empty circles.  Stars are MORPHS clusters, MORPHS [OII]-outliers are
crosses.  The dashed and dotted horizontal lines indicate the poor
group and field values, respectively.
 \label{kafrac}}
 \end{figure*}

The fractions of k+a's among all galaxies and among active galaxies
are listed for individual systems in Table~1 and shown in the left and
right panel of Fig.~\ref{kafrac}, respectively.
\footnote{The fractions considered in this section have been computed
including all spectroscopically confirmed members unweighted for
completeness.  Nothing changes if we use only a completeness corrected
sample within $R_{200}$: values remain compatible within the errors,
and the Spearman correlations presented in this section remain all
significant at more than 99\%.}
In these plots we indicate with empty symbols groups and
clusters with a low-[OII] content for their $\sigma$, i.e. the outliers in the
[OII]-$\sigma$ diagram. For comparison, we also show the values for
the MORPHS clusters, whose spectral fractions are taken from Poggianti
et al. (1999).

Both the k+a/all and -- more importantly -- the k+a/active fractions
of non-outliers appear to increase with cluster velocity dispersion.
The [OII] outliers, instead, depart from this correlation and possibly
follow a parallel one, with higher fractions at a given $\sigma$. In
fact, low-[OII] groups have a k+a incidence similar to more massive
clusters.  For EDisCS clusters the correlation probabilities are
significant (99.1\% and 99.98\%, respectively) only when [OII]
outliers are excluded. Including both EDisCS and MORPHS points, the
Spearman's test for all points ([OII] outliers and non-outliers)
yields a 99.1\% and 99.7\% probability of a correlation,
respectively. \footnote{Note that both within the EDiSCs sample and
including the MORPHS sample, the k+a/all and the k+a/active fractions
at $z=0.4-0.8$ do not correlate with redshift, indicating that in this
redshift range the dependence on cluster type dominates over the
evolution in the k+a fraction.}

\begin{figure*}
\centerline{\hspace{1cm}\includegraphics[width=0.5\columnwidth]{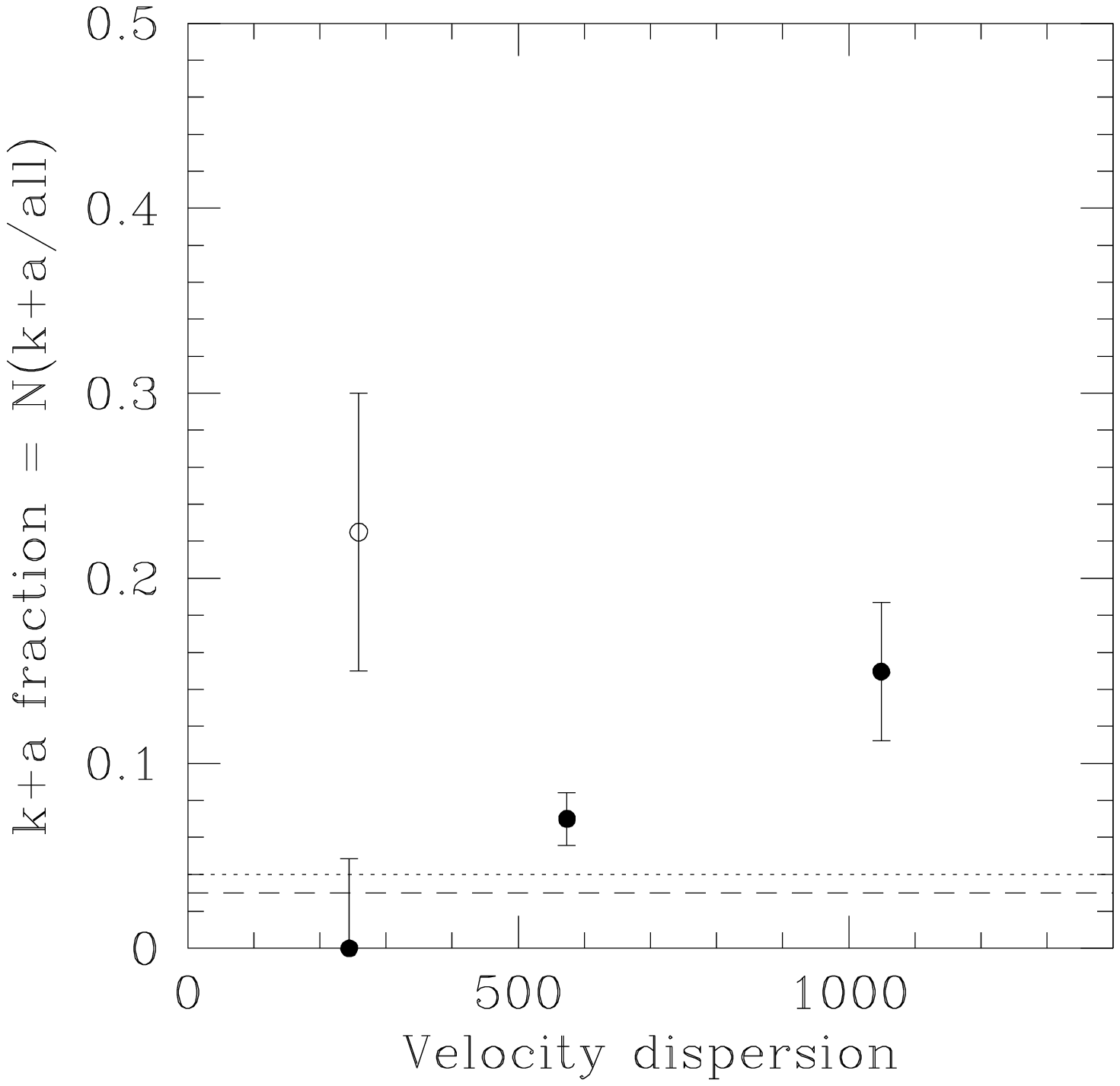}\hfill\hspace{-2cm}\includegraphics[width=0.5\columnwidth]{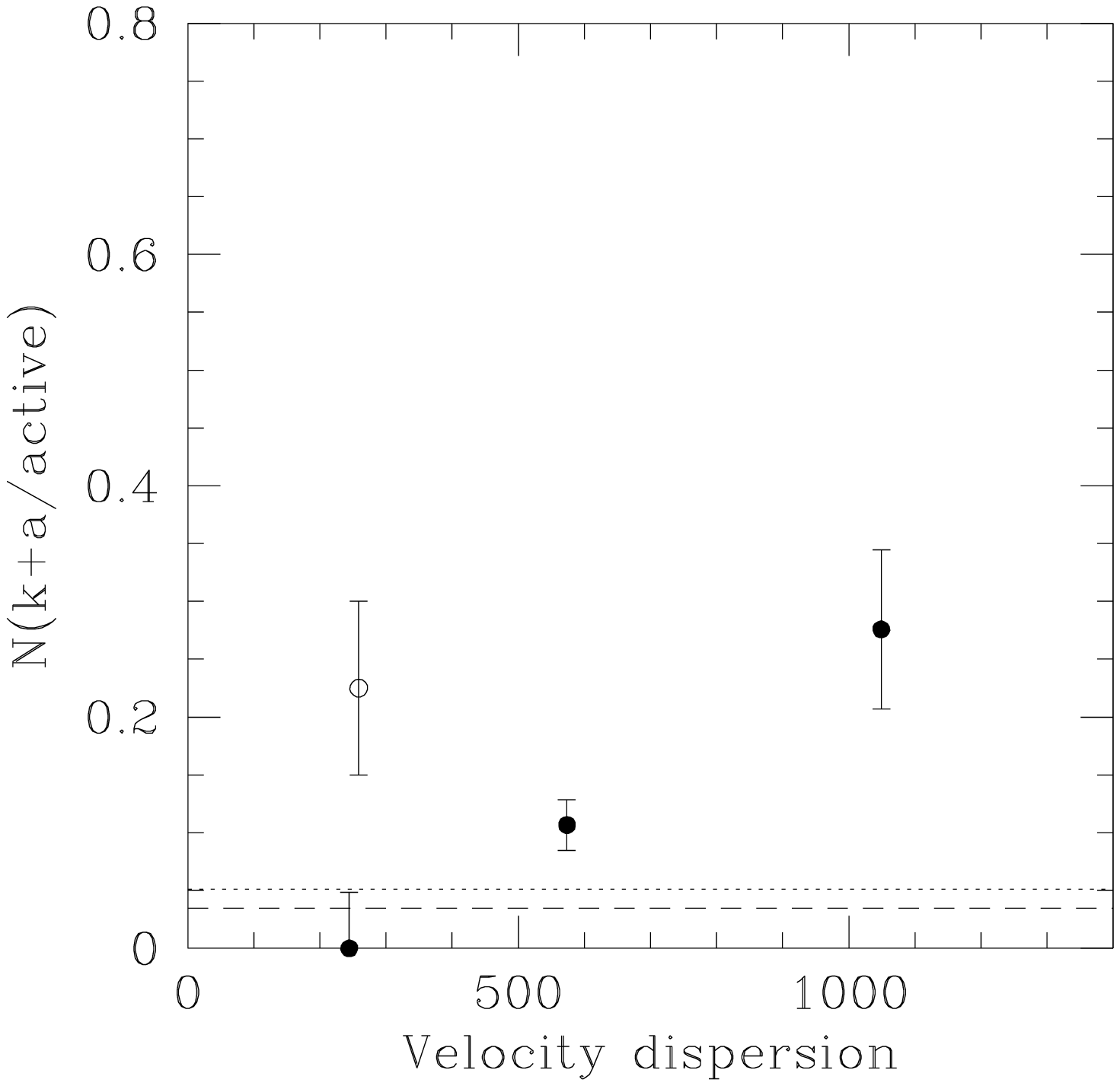}}
\caption{Same as Fig.~\ref{kafrac}, for EDisCS systems grouped in three
velocity dispersion bins. The plot shows the total fraction of k+a
galaxies among the whole cluster population (left) and among the
active population (right) for all systems with $\sigma > 750 \, \rm km \,
s^{-1}$, $400 < \sigma < 750 \, \rm km \, s^{-1}$, and for high-OII
(solid leftmost circle) and low-OII (empty circle) groups with $\sigma < 400 \,
\rm km \, s^{-1}$.  Only EDisCS structures have been included.  The
dashed and dotted horizontal lines indicate the poor group and field
values, respectively.
 \label{kafrac2}}
 \end{figure*}

A summary view of the trends with $\sigma$ is shown in
Fig.~\ref{kafrac2}, where we have grouped EDisCS structures into three
velocity dispersion bins ($\sigma > 750 \, \rm km \, s^{-1}$, $400 <
\sigma < 750 \, \rm km \, s^{-1}$ and $\sigma < 400 \, \rm km \,
s^{-1}$), keeping low- and high-[OII] groups separate (empty and solid
low-$\sigma$ points). This figure shows how, on average, the k+a
fractions in non-outliers correlate with $\sigma$. It is useful to
keep in mind that this is an average trend, while the scatter for
individual clusters can be appreciated in Fig.~\ref{kafrac}, where for
example a few intermediate-mass clusters with a null k+a fraction are
visible.  Figure~\ref{kafrac2} also visualizes the fact that, while
the average fractions in high-[OII] groups are compatible with the
field and poor group values, the other environments have average k+a
fractions higher than the field/poor groups.

\begin{figure*}
\centerline{\hspace{1cm}\includegraphics[width=0.5\columnwidth]{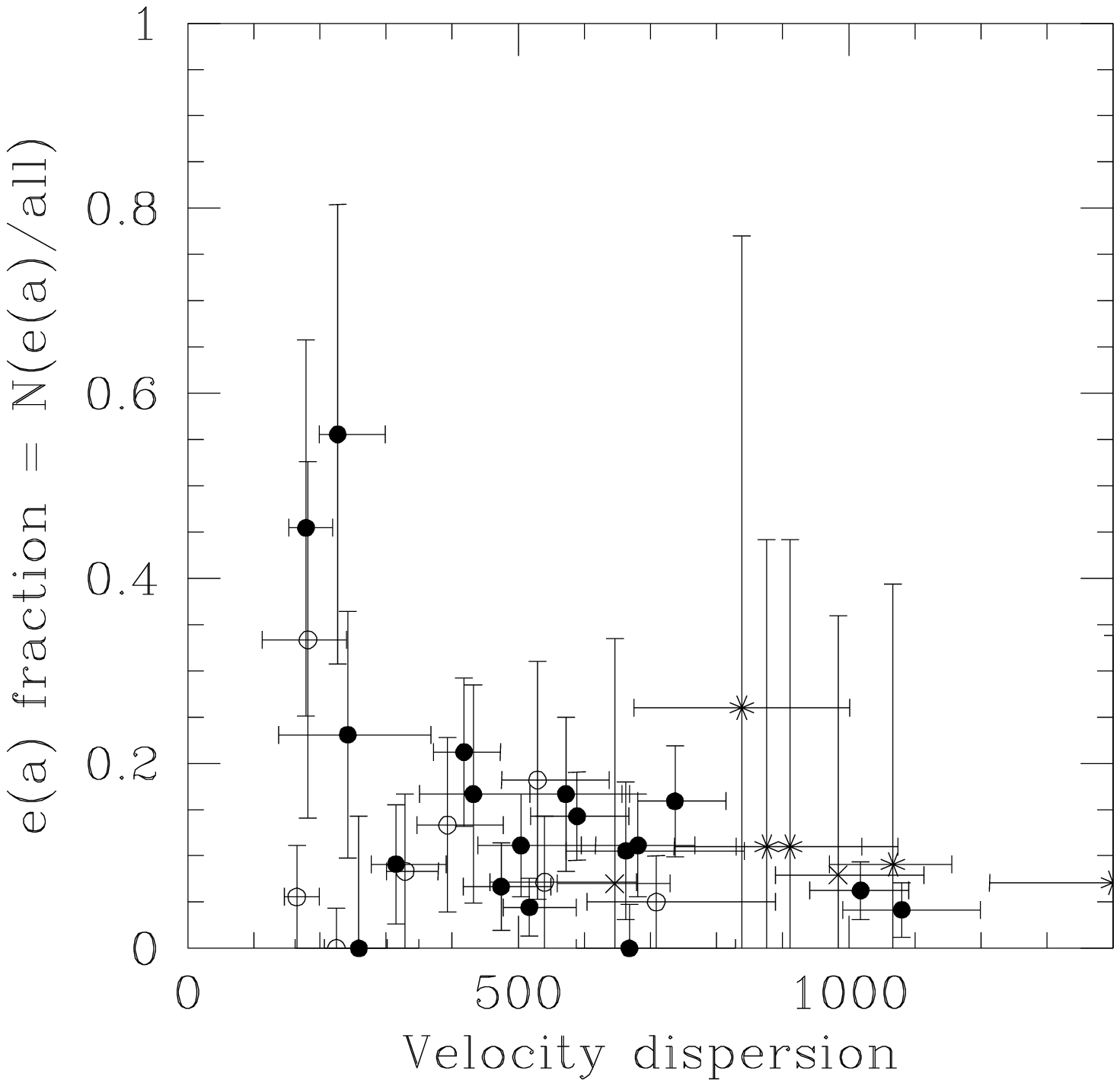}\hfill\hspace{-2cm}\includegraphics[width=0.5\columnwidth]{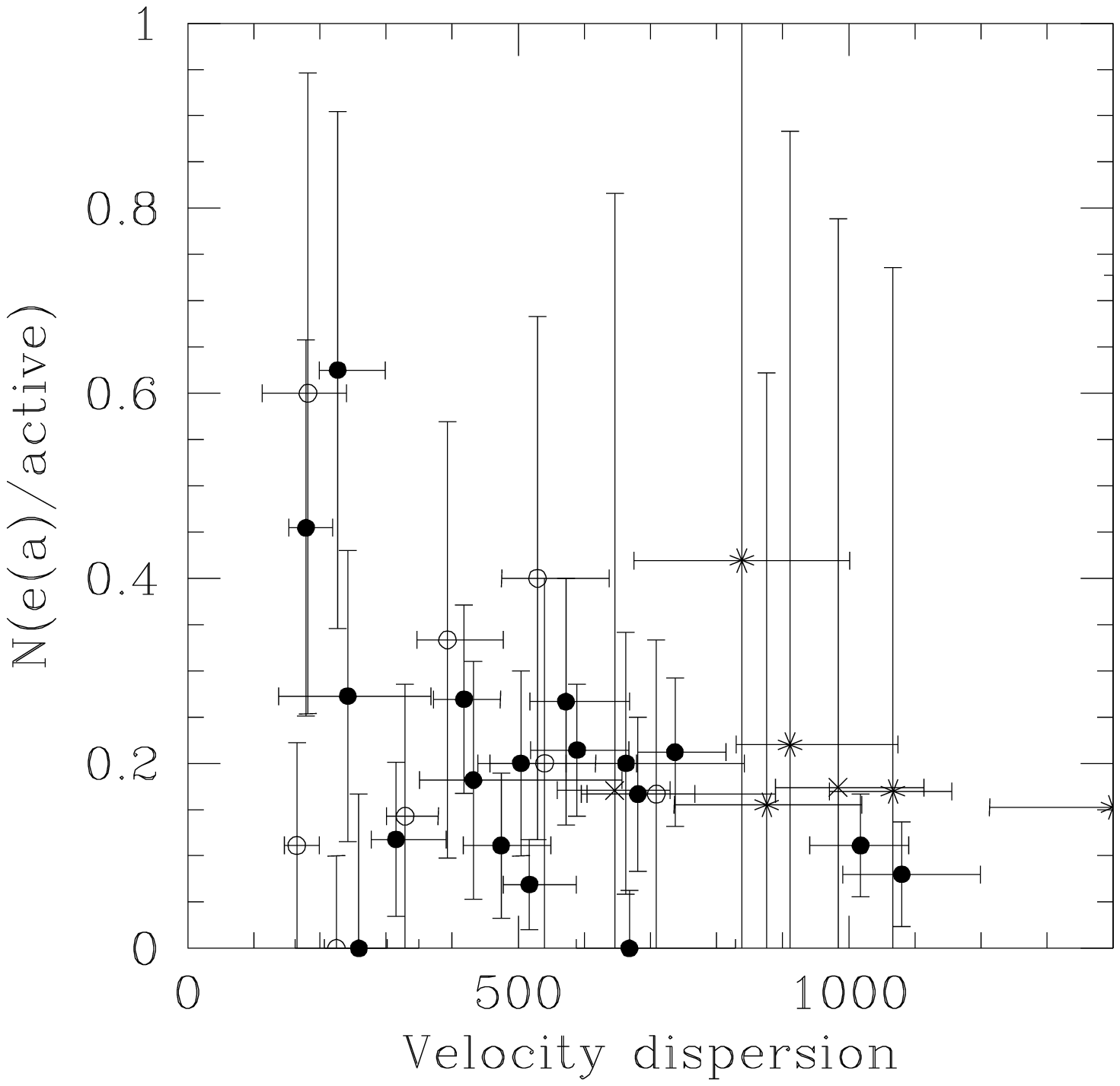}}
 \caption{Fraction of e(a) galaxies among the whole cluster population
(left) and among the active population (right) versus cluster velocity
dispersion. Circles are EDisCS structures, outliers in the
[OII]-$\sigma$ relation are indicated by empty circles. 
Stars are MORPHS clusters, MORPHS [OII]-outliers are crosses. 
 \label{eafrac}}
 \end{figure*}

In contrast to k+a's, neither the e(a)/all, nor the e(a)/active, nor
the e(a)/emission fractions, even just for non-outliers, depend on
velocity dispersion (see Fig.~\ref{eafrac}).  Hence, while the
``quenching efficiency'' depends on the cluster properties, the
incidence of dusty starburst candidates among all and among active
galaxies does not.

The trend of the post-starburst fractions with velocity dispersion is
a remarkable result, that could not be investigated in previous
studies due to the limited cluster mass range explored. It is worth
noting that the correlation of the post-starburst
fraction with cluster mass is likely to be the cause of the
differences in the global spectral fractions for EDisCS and MORPHS
clusters that were discussed in \S5 (see Fig.~\ref{kafrac} and
Table~6).  The EDisCS sample includes clusters as massive as those in
the MORPHS sample, but also a number of lower mass systems. Therefore,
the ``average'' EDisCS cluster has a lower mass than the ``average''
MORPHS cluster, and this is reflected in the lower global k+a fraction
in EDisCS clusters.

The observed trends of the k+a fractions with $\sigma$ raise two
main questions:

1) Why are more massive systems progressively more efficient in quenching
star formation  (the k+a/active fraction).

2) What physical conditions and/or structure history set the outliers
apart from the other systems, and cause them to quench star formation
with a higher efficiency than other systems of their mass.

These questions will be addressed in \S8, after having gathered
additional observational clues.

\begin{figure*}
\centerline{\hspace{1cm}\includegraphics[width=0.5\columnwidth]{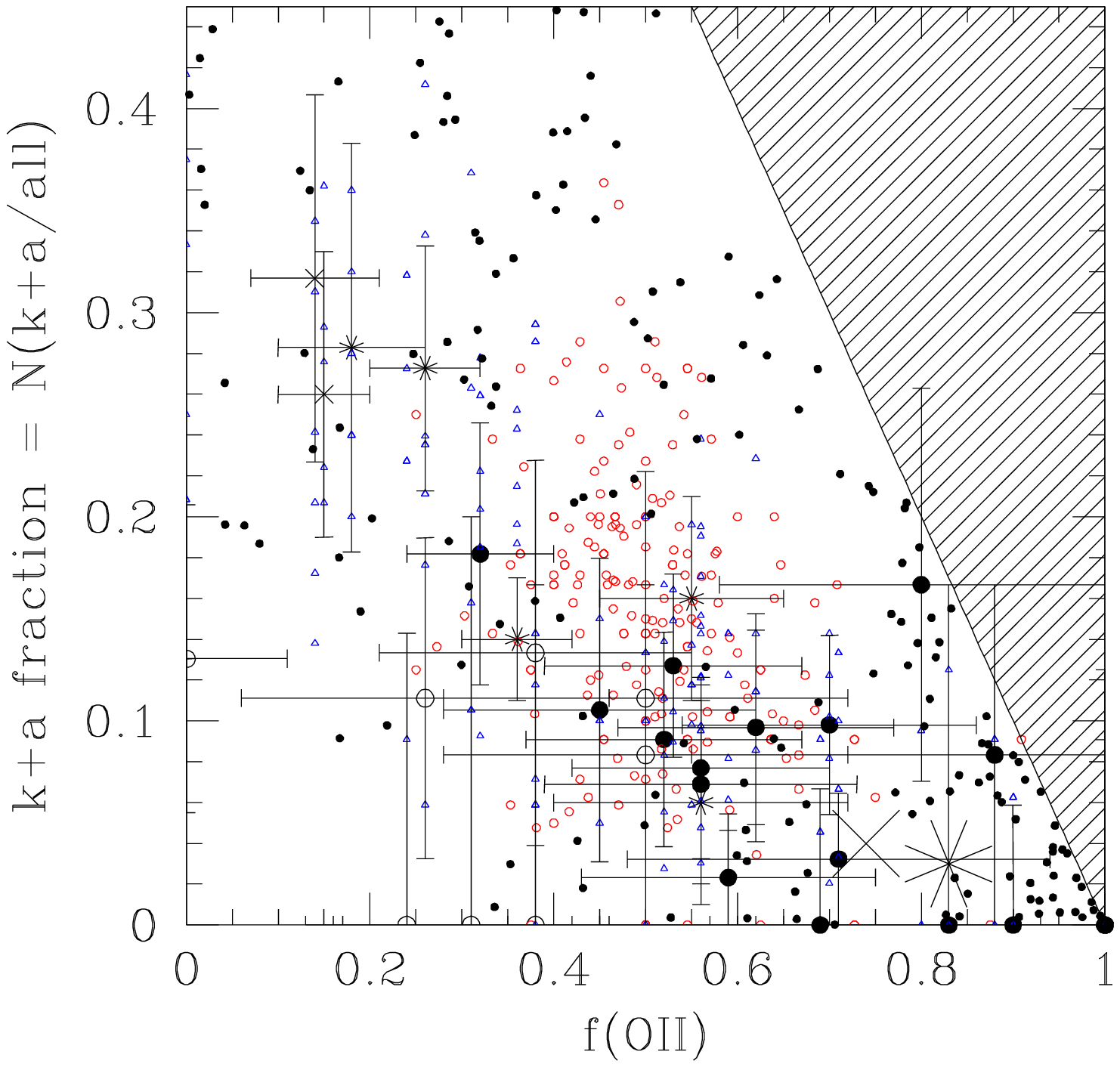}\hfill\hspace{-2cm}\includegraphics[width=0.5\columnwidth]{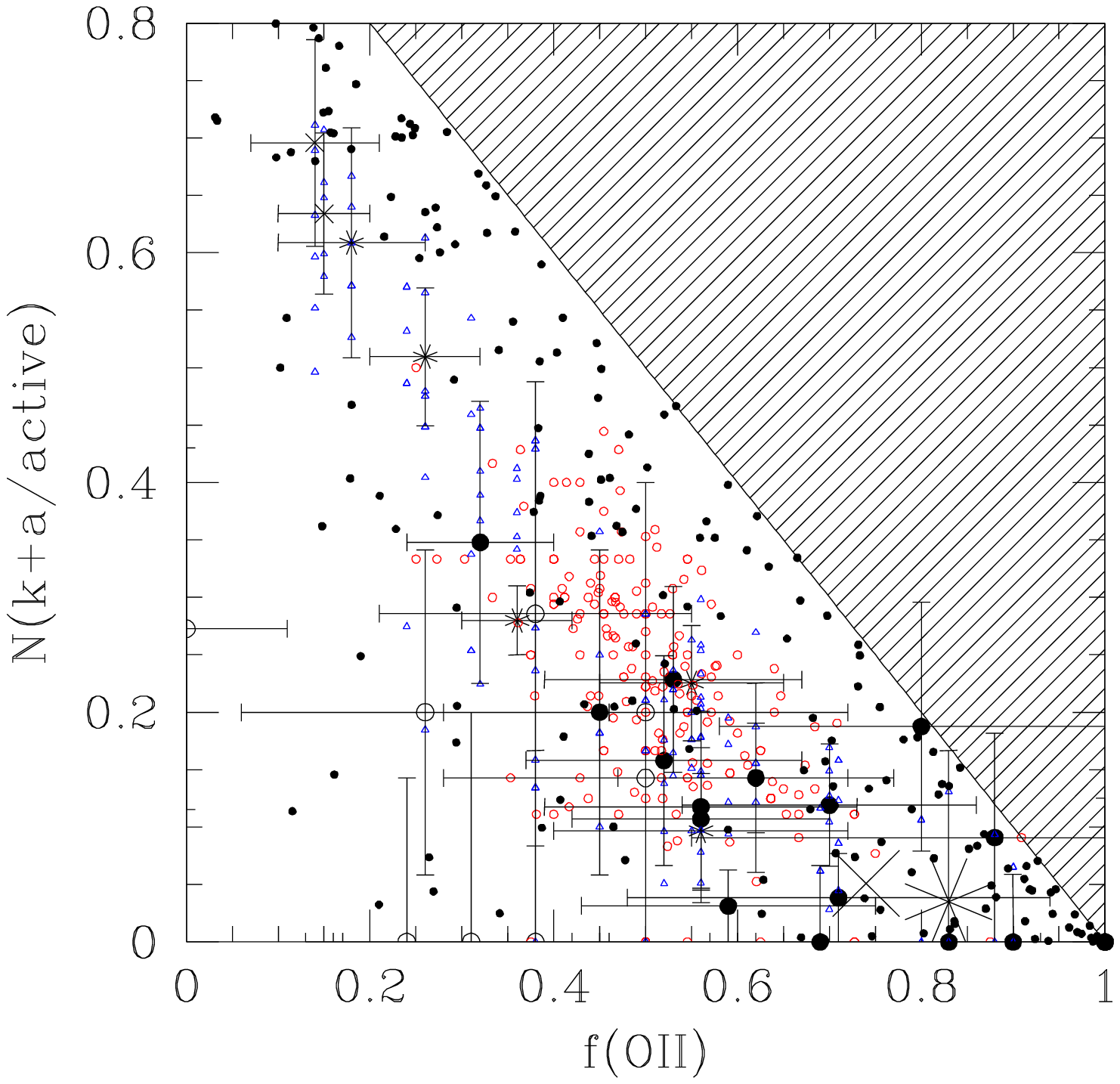}}
 \caption{Fraction of k+a galaxies among the whole cluster population
(left) and among the active population (right) versus fraction of
[OII] emitters. The large star and cross indicate the poor group and
field values, respectively. Other symbols as in Fig.\ref{kafrac}.  The
shaded area is the ``zone of avoidance'' that cannot be populated,
given the definition of the X and Y axes (see \S5.2).  Small black
points represent the location of 200 simulated systems with a random
f([OII]) and a random k+a fraction (first test in \S5.2).  Red empty
circles are the results for a binomial $f_{OII}$ probability
distribution centered around 0.5 and adopting a 30\% probability that
any one of the non-star-forming galaxies is a k+a (second test in
\S5.2). Blue empty triangles are the results adopting the observed
distribution of $f_{OII}$ and assuming a 30\% probability that any one
of the non-star-forming galaxies is a k+a (third test in \S5.2).
 \label{kaoii}}
 \end{figure*}

We proceed comparing the k+a fractions with the star-forming fraction
in Fig.~\ref{kaoii}, where the latter includes galaxies with
EW([OII])$>3$ \AA $\,$ as in Poggianti et al. (2006). 
Both the k+a/all and k+a/active fractions 
anticorrelate with the fraction of [OII] emitters (96.4\% and
99.1\% in EDisCS, 99.98\% and 99.91\% if both EDisCS and MORPHS are
included). The peculiarity of most outliers disappears in this
diagram, because they align with the rest of the systems.

We note that, at a fixed [OII] fraction $f_{[OII]}$, both the k+a/all
and the k+a/active fractions can be at most equal to (1-$f_{[OII]}$).
As a consequence, in Fig.~\ref{kaoii} there are ``zones of
avoidance'', shown as shaded regions, that cannot be populated by
datapoints due simply to the definition of the quantities on the X and
Y axes. However, this upper limit cannot induce forced
anticorrelations as those we observe. 

To further investigate whether the anticorrelation could arise as a
result of the definition of the different galaxy classes, we perform
two types of simulations, simulating 200 clusters with a distribution
of number of member galaxies equal to that observed.  For the first
simulation, we show in Fig.~\ref{kaoii} the position of the simulated
systems starting from a uniform distribution of f([OII])'s between 0
and 1 and allowing the k+a fraction to vary randomly and uniformly
between 0 and (1-f([OII])) (black small circles).  The simulated
systems are scattered all over in the k+a/all vs f([OII]) diagram,
except for the avoidance zone.  In the k+a/active vs f([OII]) diagram,
most points of the random distribution are concentrated in a stripe
parallel and close to the avoidance limit.  In both cases, the
observed anticorrelations are incompatible with the distribution
expected from the simulation, as confirmed by a 2D Kolmogorov-Smirnov
test.


In the second test, in all of the simulated clusters we draw random
realizations assuming binomial distributions with probabilities of the
galaxies being star-forming equal to 50\%, and adopting a 30\%
probability that any one of the non-star-forming galaxies is a k+a.
This test is shown as the open red circles in Fig.~\ref{kaoii}.  The
cloud of simulated clusters illustrate something akin to a ``point
spread function'' in these two diagrams.  As expected, the simulated
points cluster around $f_{OII}=0.5$ and $f_{k+a}=0.15$. They display a
very large scatter in k+a/all fraction, failing to account for the
observed trend. A 2-dimensional Kolmogorov-Smirnov test rules out the
hypothesis that the observed and the simulated distributions are drawn
from the same parent distribution.

These simple tests show that the mere definitions of the plotted
quantities is not sufficient to explain the observed distribution of
points, suggesting a physical origin for the link between the
frequency of the various spectral classes.

To investigate such an origin, we ran a third simulation, starting
from the observed distribution of $f_{OII}$ as measured by Poggianti
et al. (2006). In this work, the $f_{OII}$ fraction was found to
follow a general broad anticorrelation with cluster velocity
dispersion, although with a number of outliers.  For the galaxies
determined to be non-emission line galaxies in each cluster we then
randomly determine whether they are k+a galaxies, adopting a 30\%
probability that any one of the non-emission line galaxies is a k+a.
Results are shown as blue open triangles in Fig.~\ref{kaoii} and are
compatible with the trends observed in both panels.  For this third
case, Kolmogorov-Smirnov tests allow the observed and the simulated
distributions to be drawn from the same parent distribution (the KS
probabilities for the distributions {\it not} to be drawn from the
same parent distribution are 0.109 and 0.703).

Assuming a constant k+a probability {\it among non-emission-line
galaxies} (0.3 in the third simulation) physically means a higher
k+a production rate {\it per emission-line galaxy} in systems with
lower star-forming fractions. In fact, it is important to keep in mind
that the population of emission-line galaxies is the ``reservoir'' to
produce k+a's, thus a fixed k+a probability among non-emission-line
galaxies corresponds to adopting a higher conversion efficiency from
emission-line to k+a in systems with lower star-forming fractions.

Therefore, the anticorrelation in Fig.~\ref{kaoii} might point to a
connection between the ``quenching efficiency'' and the fraction of
cluster galaxies that still retain an ongoing star formation activity
or, equivalently, the fraction of cluster galaxies that are devoid of
star formation (1-f([OII]). Naively, this connection seems to imply
that the observed proportion of passively evolving galaxies in a
system depends directly on the capacity of the cluster/group to
truncate the star formation activity in currently infalling galaxies.
Assuming that the SF activity, once quenched, does not re-start at
later times, this result is perhaps unsurprising if environments that
are good at quenching at the moment we observe them have been good at
quenching also at previous times, even if the quenching mechanism may
not be the same over time.  Attempting an interpretation of the
observed anticorrelation, however, requires an articulated model for
the accretion and quenching history of clusters, that is beyond the
scope of this paper.

\section{Composite spectra}

Composite cluster spectra can be obtained combining the light of
individual galaxies.  This method is complementary to the analysis of
single galaxy spectral fractions, and provides a measurement of the
``global'' star formation history integrated over all galaxies in a
cluster (Dressler et al. 2004, hereafter D04).  The high
signal-to-noise ratio of composite spectra allows a robust comparison
of the global stellar history as a function of environment and
redshift.

A composite spectrum for each of our clusters was obtained by summing
the spectra of cluster members, 
after normalizing each spectrum by its
mode and assigning a weight equal to its observed I-band luminosity. 
Each spectrum was further weighted for spectroscopic incompleteness as a
function of galaxy magnitude and position in the cluster
(see \S2).\footnote{Measuring the [OII] and $H\delta$ equivalent widths
on both weighted and unweighted composite spectra, we found that
weighting the composites for spectroscopic incompleteness does not
change any of the results.} The EW([OII]) and EW($\rm H\delta$) for
our composite spectra are listed in Table~1.

\begin{figure}
\centerline{\hspace{2cm}\includegraphics[width=8cm]{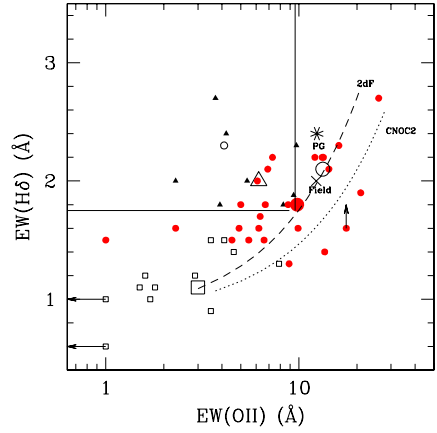}}
\caption{Rest frame equivalent widths of [OII] and $H\delta$ from
composite spectra for 3 datasets of clusters: Dressler \& Shectman low
redshift clusters as in D04 (empty squares); $z=0.4-0.5$ MORPHS clusters
as in D04 (filled triangles) and EDisCS structures at $z=0.4-0.8$ (filled
circles).  The average of each dataset is indicated by {\it large}
symbols: empty square for low-z clusters, empty triangle for MORPHS
and filled circle for EDisCS.  An additional rich cluster at $z\sim 0.8$
(MS1054-03 from D04) is indicated by the small empty circle.  The
values for EDisCS poor group, field and low-OII group composites are
represented by the asterisk, the cross and the large empty circle,
respectively. The dashed and dotted lines describe the locus of mixes
of passive and continuously star-forming galaxies for low-z (2dF) and
$z \sim 0.4$ (CNOC2) field galaxies, respectively, taken from D04.
The vertical and horizontal lines are drawn to highlight the region
occupied by MORPHS clusters and corresponding to a Balmer excess.
 \label{d04}}
 \end{figure}

As suggested by D04, the most useful diagnostic diagram is a plot of
EW($\rm H\delta$) versus EW([OII]) (Fig.~\ref{d04}). In this diagram,
clusters with a Balmer-strong (e.g. a prominent post-starburst)
population will lie at higher EW($\rm H\delta$) for a given EW([OII]) than
clusters without a Balmer enhancement.  The local reference location
in this plot is taken from D04: low-z clusters are shown as empty
squares and have low-[OII] and low $\rm H\delta$.  The expected
location of the ``normal'' field population is shown by the dashed and
dotted lines, also taken from D04. These represent mixes of different
proportions of passive and quiescently star-forming galaxies, which are
supposed to
represent the field at different z's and to follow the variation in
the star-forming fraction with redshift.  MORPHS clusters at
$z=0.4-0.5$ (filled triangles) show an $\rm H\delta$ excess and occupy
a region of the diagram separated from the location of both low-z
clusters and the evolving field.  We identify the region occupied by
MORPHS clusters as a box in the upper left corner of Fig.~\ref{d04}.

EDisCS clusters and groups are shown as solid circles.  Only a few of
them lie in in the upper left box where MORPHS clusters lie.  These
are namely Cl\,1103a, Cl\,1119, Cl\,1054-12a, and, marginally,
Cl\,1216, Cl\,1420 and Cl\,1354.  The great majority of the low-[OII]
groups are therefore found inside the $H\delta$-excess box.  Other
$H\delta$-excess systems might be Cl\,1138a and
Cl\,1232.\footnote{Measurements for Cl\,1232 should be taken with
caution, because they reach out only to 1/2 $R_{200}$ and because the
6300 sky line falls on top of $H\delta$ at this redshift.}  The [OII]
and $H\delta$ EWs of the composite spectrum of the $H\delta$-excess
systems (named ``Excess clusters and groups'') are given in Table~1,
and coincide precisely with those in the MORPHS cluster composite, taken
from D04 and shown as large empty triangle.

In contrast, the majority of our other clusters and groups display no
clear $\rm H\delta$ excess, being located around the region of the
diagram corresponding to combinations of ``normal''
star-forming and passively evolving galaxy spectra. Composite 
high-OII groups, poor groups, field and all our cluster and groups
together (large empty circle, asterisk, cross, large filled circle,
respectively) all lie close to the ``normal galaxy'' sequence.

The analysis of cluster composite spectra thus reinforces the
conclusions based on individual galaxy spectra and presented in the
previous section: only a subset of the distant clusters and groups
have an outstanding Balmer strong population.  The latter appears to
be conspicuous in low-[OII] groups and some clusters, that tend to be
mainly the most massive ones.
The rest of our clusters and groups, as well
as the poor groups and the field, do not show a notable
$H\delta$-excess.

\section{Properties of starburst, post-starburst, quiescently and passively evolving galaxies}

In the following, we analyze other galaxy properties and the local
environment of galaxies of different spectral types, that may shed
some light on the origin of the various star formation histories and
the evolutionary link between them.

\subsection{Morphologies}

\begin{figure}
\centerline{\hspace{2cm}\includegraphics[width=10cm]{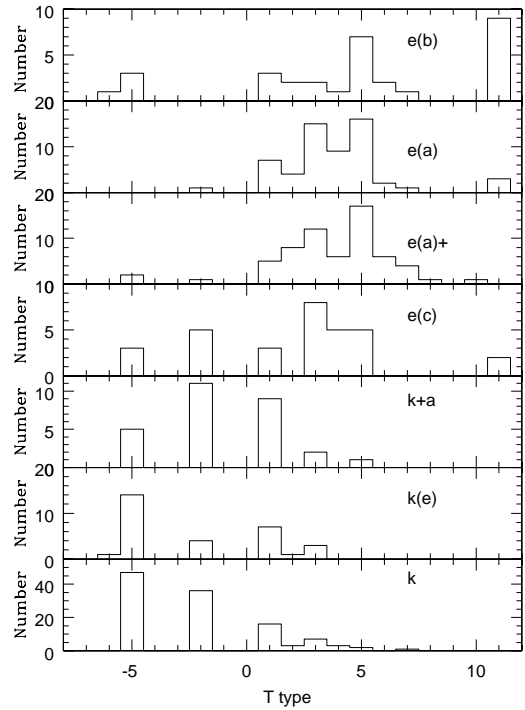}}
\caption{Morphologies of galaxies of different spectral types.
 Nonstellar but too compact to see structure = -6, E=-5, S0=-2, Sa=1,
 Sb=3, Sc=5, Sd=7, Sm=9 Irr=11.
 \label{morphs}}
 \end{figure}

Visual morphologies based on ACS/HST images are available for 10 of
the 20 EDisCS fields (Desai et al. 2007).  The morphological
distribution of galaxies of different spectral types is shown in
Fig.~\ref{morphs}.  The majority of passively evolving (k) and k(e)
galaxies are ellipticals and S0's, with a small tail of Sa's and later
types.  Post-starburst (k+a) galaxies are mostly early disk galaxies
of types S0's and Sa's.  Actively star-forming galaxies, as expected,
are mostly spirals, with e(c)'s being a mix of Sb's, Sc's and some
earlier types, e(a)'s and e(a)+'s being typically Sb and Sc galaxies,
and e(b)'s tending to have on average later type morphologies,
principally Sc's to Irregulars. Most irregular galaxies have an e(b)
spectrum, as in the local Universe.

In general, the morphological mix follows what is expected from the star
formation properties, with early-type, passively evolving galaxies and
late-type star-forming galaxies, and k+a galaxies that are transition
objects with early disk morphologies. The k+a morphological
distribution we observe is shifted towards slighly earlier
morphologies than that observed by the MORPHS (Dressler et al. 1999).
Our k+a's are typically S0's and Sa's, while several MORPHS k+a's
display later spiral morphologies.

\subsection{Masses, luminosities and mass-to-light ratios}

\begin{figure*}
\centerline{\hspace{1cm}\includegraphics[width=8cm]{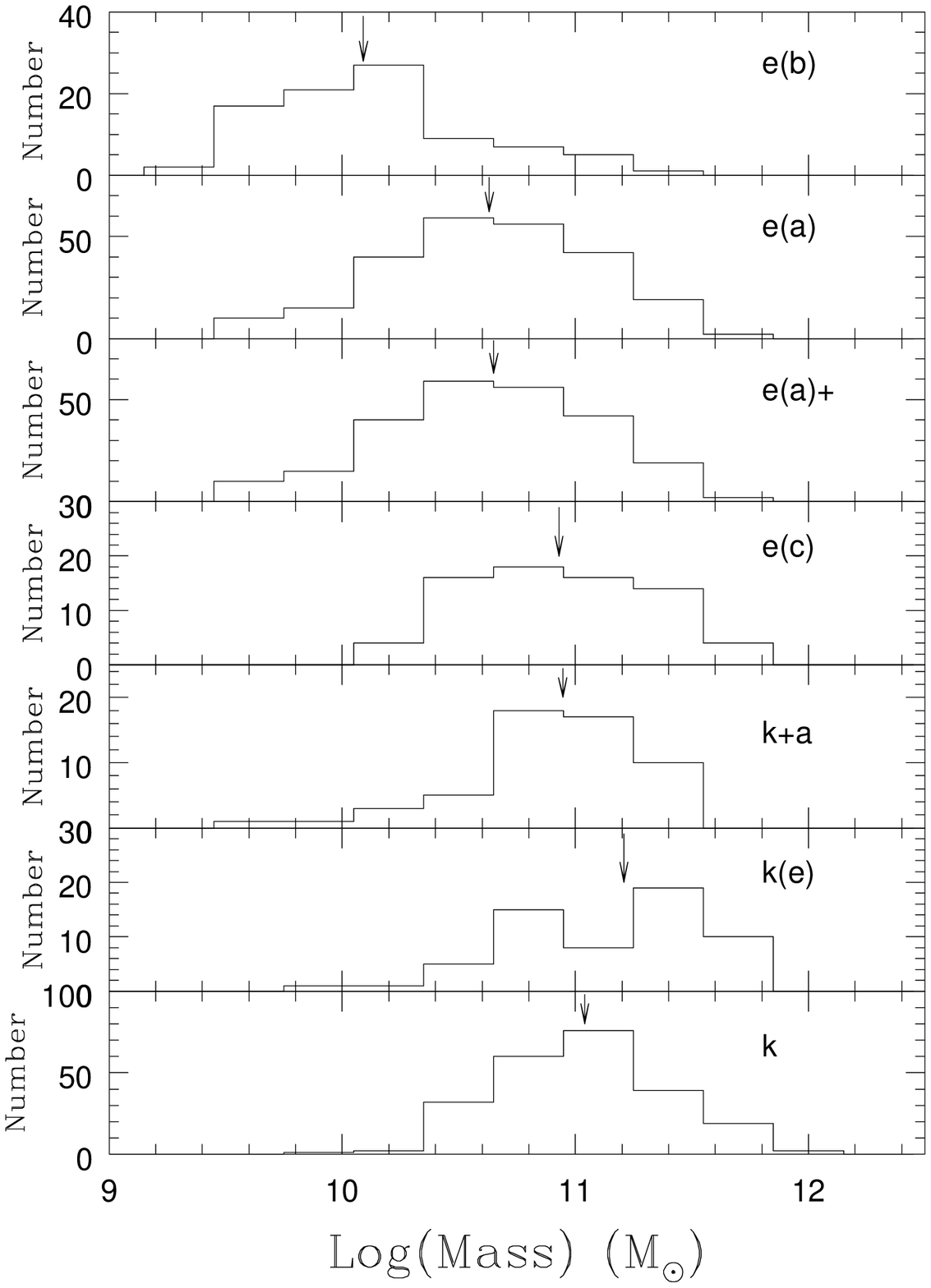}\hfill\hspace{-2cm}\includegraphics[width=8cm]{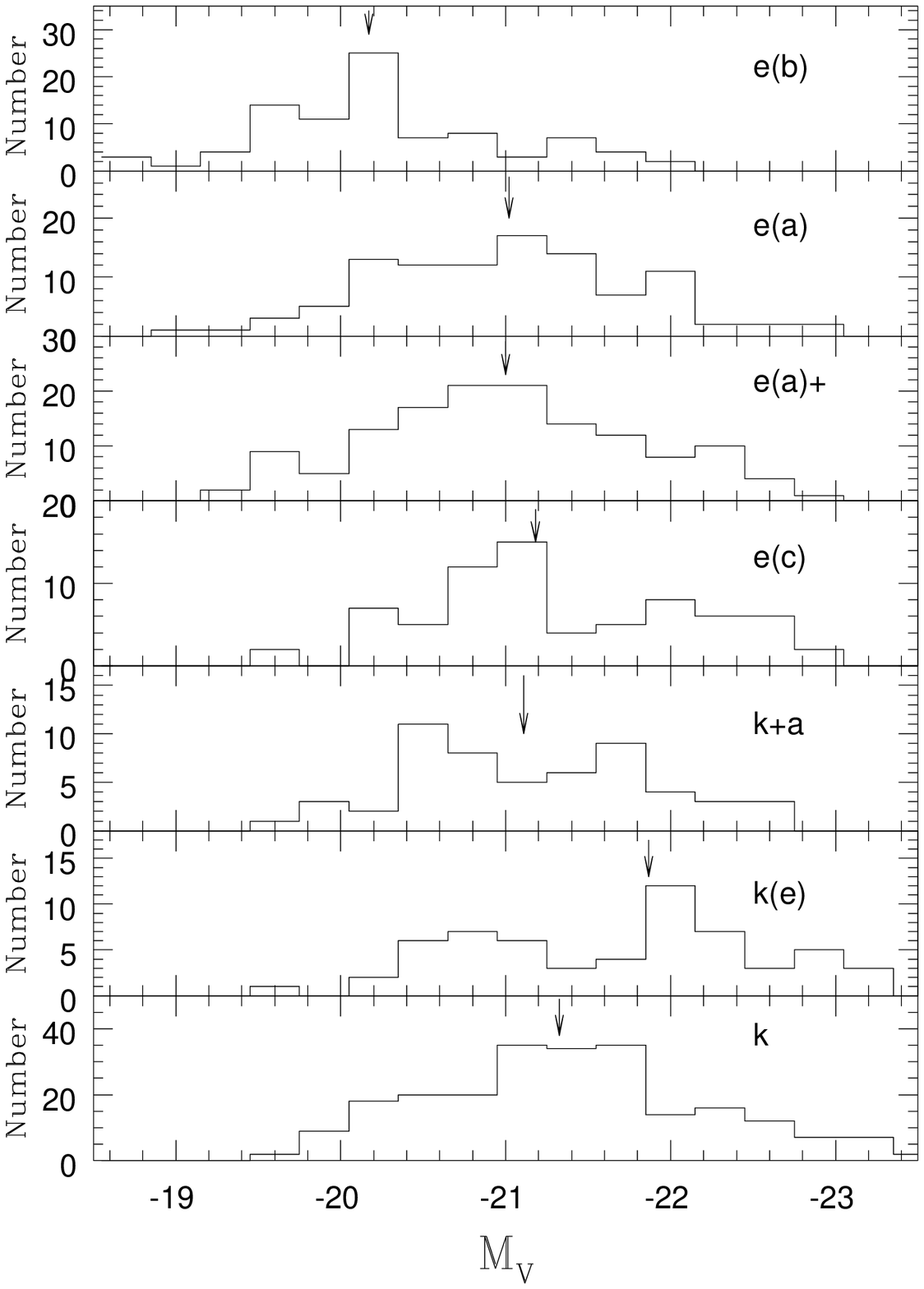}\hfill\hspace{-2cm}\includegraphics[width=8cm]{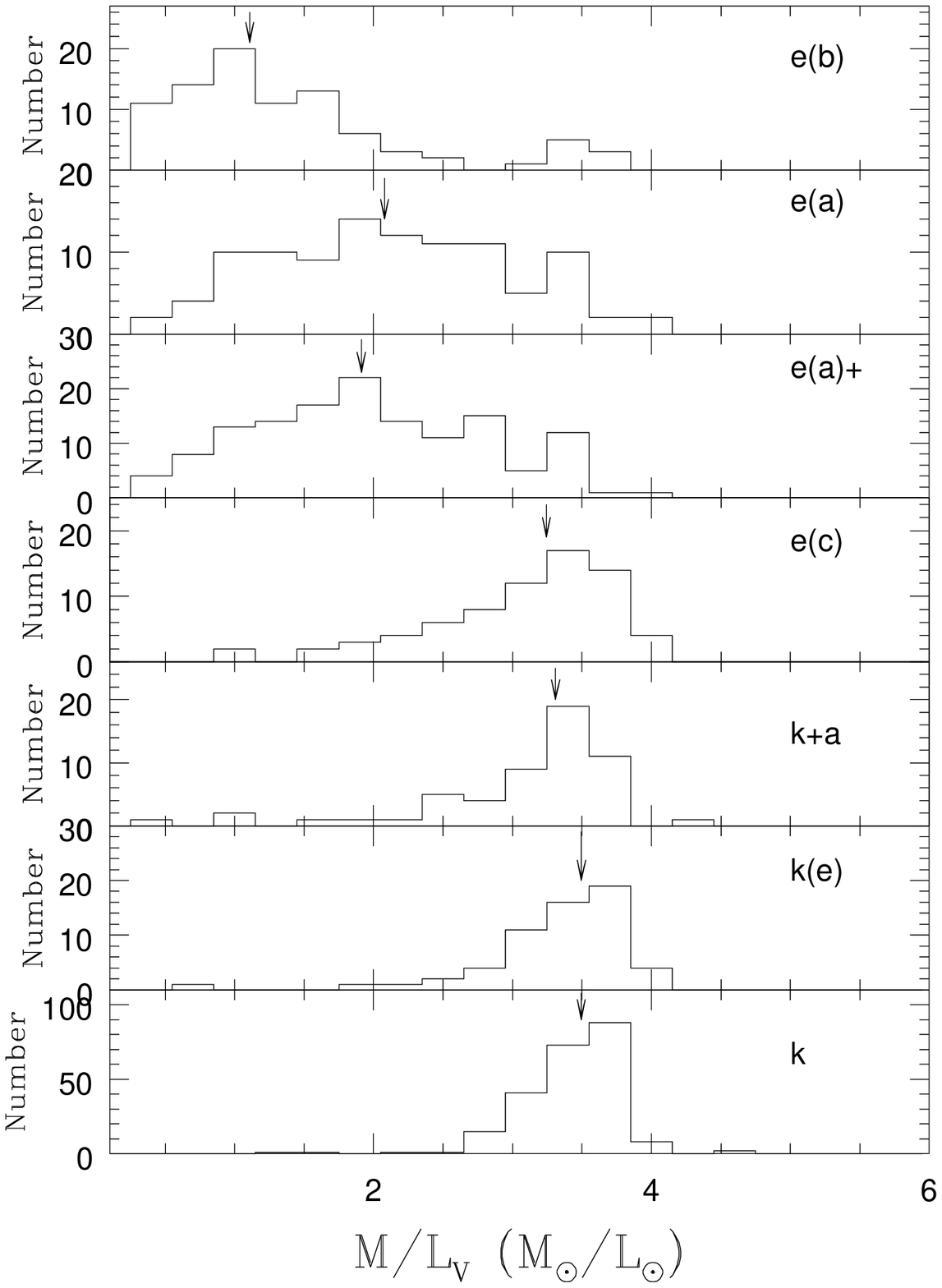}}
 \caption{Mass (left), absolute V magnitude (center) and Mass-to-Light ratio
(right) distributions of galaxies of different spectral types. The 
arrows indicate the median value for each type.
 \label{masses}}
 \end{figure*}

Stellar mass, absolute V magnitude and 
Stellar Mass-to-Light ratio distributions of
galaxies of different spectral types are shown in Fig.~\ref{masses}.

As it is natural to expect on the basis of what has been shown above
and in previous works, the spectral classification is also a sequence
of decreasing average galaxy mass and decreasing mass-to-light ratio
going from k/k(e), k+a, e(c), e(a)/e(a)+ to e(b) types.  The absolute
magnitude distribution, instead, is more uniform across spectral
types, except for e(b) galaxies that are the least massive, but also
the least luminous, class.

This strengthens some of the conclusions found by previous works,
and provides some new insights:

a) strong emission-line (e(b)) galaxies are not the primary
progenitors of k+a galaxies, being significantly less massive than
k+a's (Poggianti et al. 1999). 
The obvious candidate progenitors of the bright/massive
k+a's included in our sample are star-forming galaxies of the e(a), e(a)+
and e(c) types.

b) since the stellar mass distributions of k and k+a galaxies are
rather similar, the fate of the observed k+a's (assuming they never
resume a star formation activity) is to increment the class of
massive, passively evolving k galaxies.  Given also their morphologies
(\S7.1), k+a's are massive galaxies in the transition phase from
actively starbursting or star-forming disk galaxies to passively
evolving early-type galaxies.  This scenario has been proposed and
discussed before in several works (e.g. Poggianti et al. 1999), 
although galaxy masses were not available then.

c) if this scenario can explain the evolution with redshift of galaxy
properties (star formation activity and morphologies) in clusters and,
as we now find, in the low-[OII] groups, the paucity or even lack of
k+a galaxies in other environments leads to the question of whether/how
galaxies turn from blue actively star-forming to red passively
evolving in high-[OII] groups, poor groups and the field.  Currently
it is unknown how to what extent the observed
strong evolution of the number ratio
of red to blue galaxies on a cosmic scale (e.g. Bell et al. 2007)
can be accounted for by the
presence of k+a galaxies in the different environments and,
conversely, how many galaxies in each environment turn from blue to
red without going through a k+a phase.  It is hard to assess 
quantitatively
the impact of k+a galaxies on the cosmic star formation, 
not knowing the relative proportion of low-[OII] and
high-[OII] groups, and what fraction of all galaxies could be
processed through the low-[OII] group environment.

\subsection{Local densities and radial distributions}

The location of galaxies of different spectral types within the
cluster may provide some clues about the physical cause for the spectral
characteristics and, therefore, the star formation pattern.

\begin{figure}
\centerline{\hspace{2cm}\includegraphics[width=10cm]{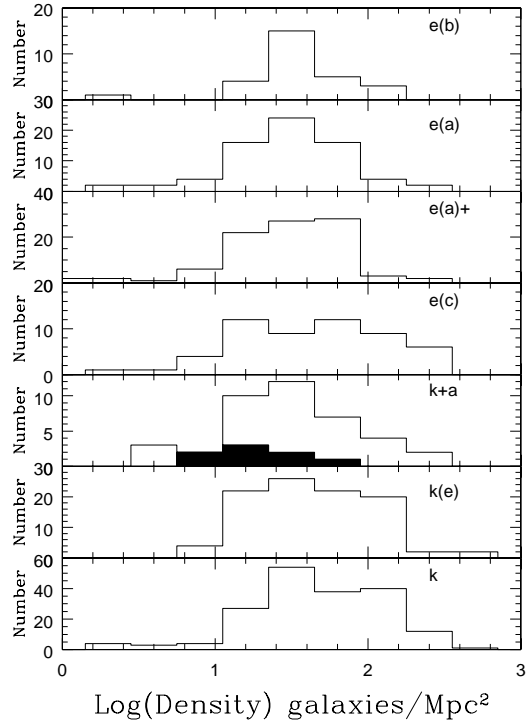}}
\caption{Projected local density distributions of galaxies of each
 spectral type. The filled histogram in the k+a panel represent
the youngest k+a's (see text).
 \label{den}}
 \end{figure}

Distributions of projected local density for galaxies in our clusters
and groups are shown in Fig.\ref{den} for each spectral class. The
local galaxy density was computed within the circular area enclosing
the closest 10 neighbours, as described in Poggianti et
al. (2008). E(a), e(a)+ and e(b) galaxies are rare in regions of
densities higher than 100 galaxies/$\rm Mpc^2$.  The same is true for
the youngest k+a galaxies (solid histogram), identified as those with
the strongest Balmer lines and blue colors (r.f.  $B-V \leq
0.7$, $EW(\rm H\delta)>4\AA$ and $EW(\rm H\zeta)>5\AA$), that tend to
inhabit, on average, lower density regions than the general k+a
population.

\begin{figure*}
\centerline{\hspace{2cm}\includegraphics[width=10cm]{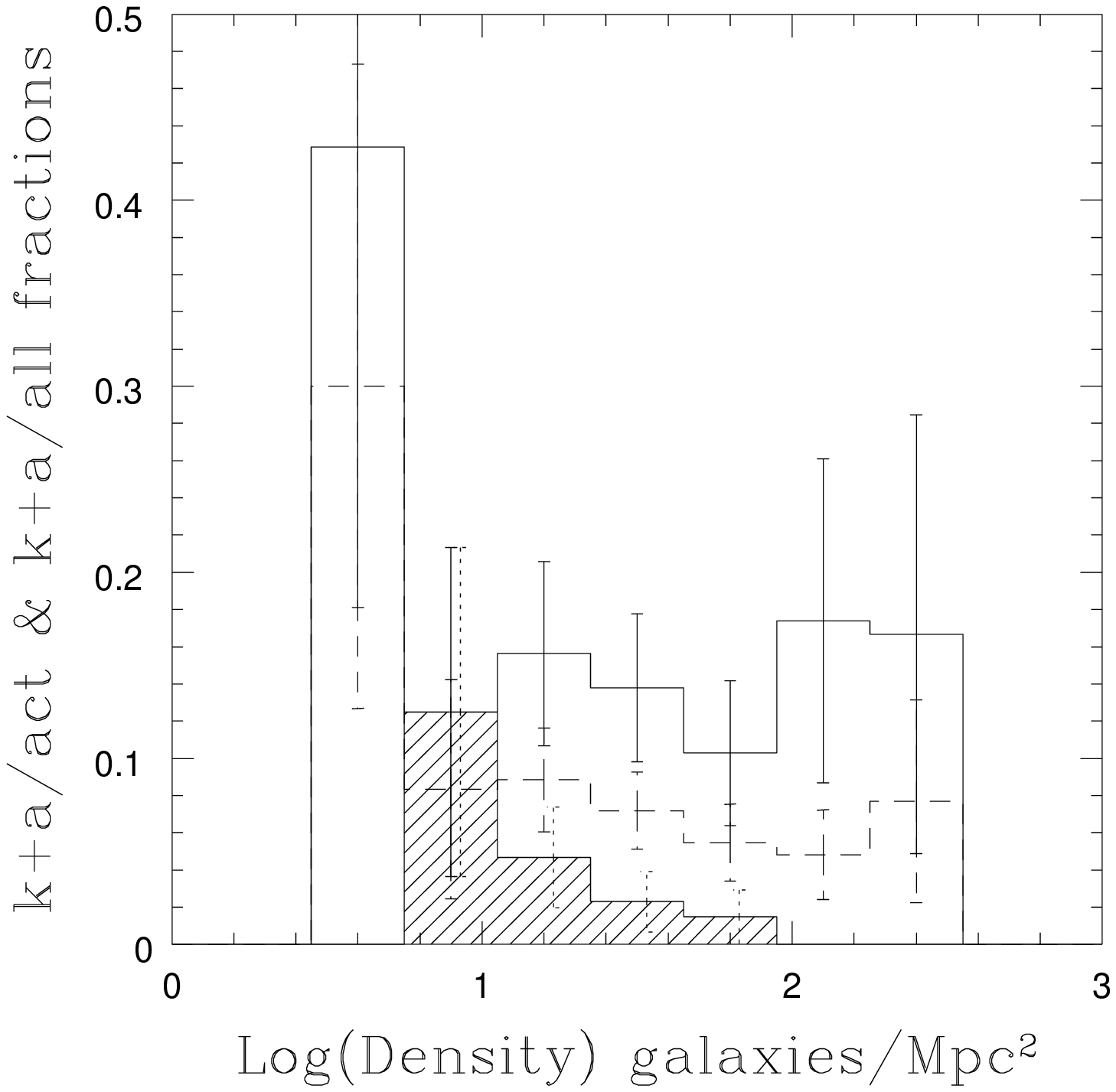}\hfill\hspace{-2cm}\includegraphics[width=10cm]{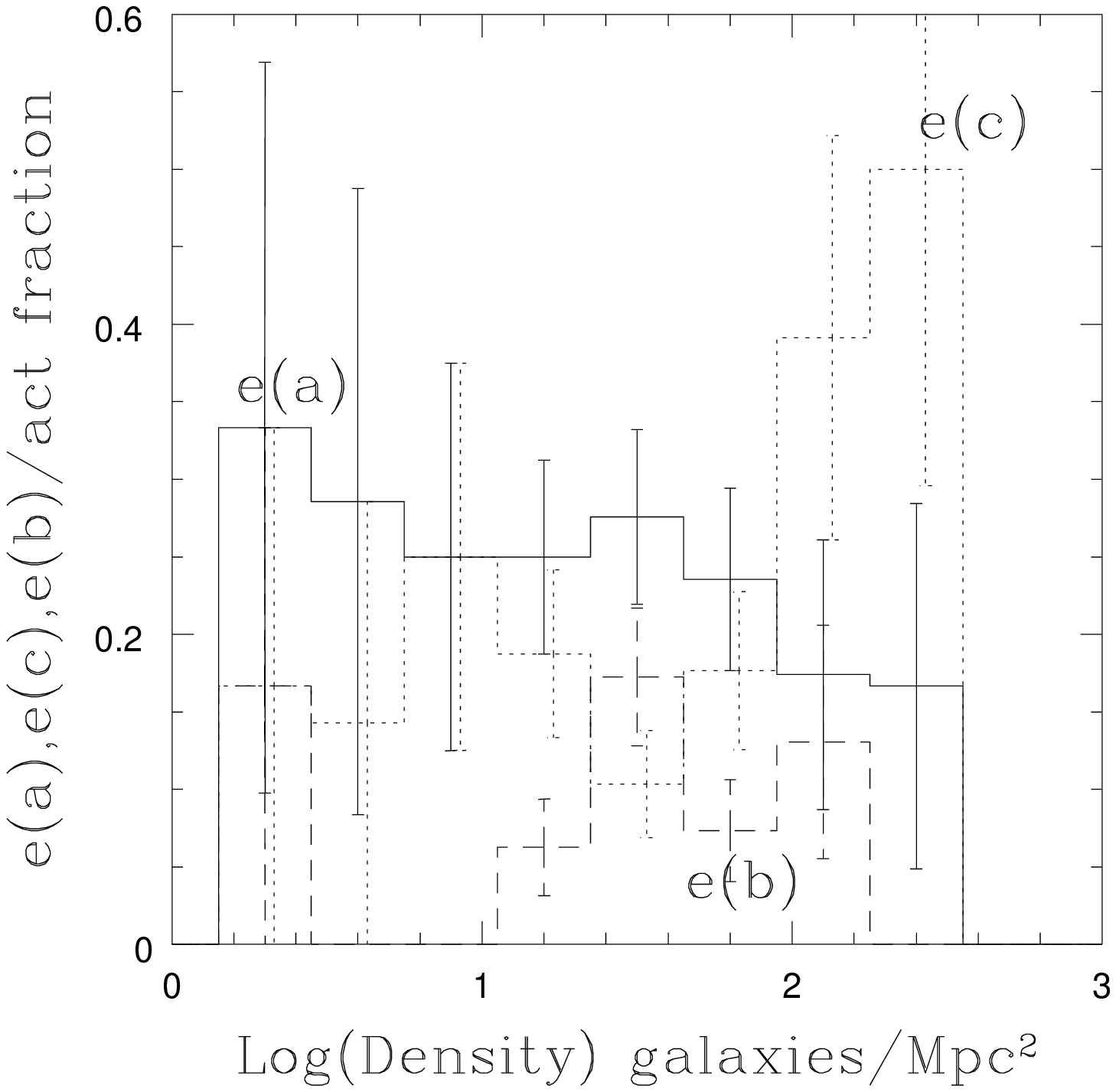}\hfill}
 \caption{{\bf Left} Fraction of k+a galaxies among active galaxies
(k+a/act) as a function of local density for EDisCS cluster and group
galaxies. Solid empty histogram and solid errorbars: all k+a's.
Shaded histogram and dotted errorbars: young k+a's (see text). Dashed
histogram and dashed errorbars: k+a fraction among all galaxies
(k+a/all). Errorbars are computed from Poissonian statistics.  {\bf
Right} Fraction of e(a) (solid histogram and errorbars), e(c) (dotted
histogram and errorbars) and e(b) (dashed histogram and errorbars)
galaxies among active galaxies as a function of local density for
EDisCS cluster and group galaxies.
 \label{den2}}
 \end{figure*}

Even more significant is to look at the incidence of post-starbursts
and dusty starburst candidates among the active population as a
function of local density (Fig.~\ref{den2}).  The overall k+a/active
and k+a/all fractions are rather flat with density, except for a
possible peak at the lowest densities we sample. Isolating the
youngest k+a's (shaded histogram in the left panel of
Fig.~\ref{den2}), their incidence tends to {\it increase} towards {\it
lower} densities. A similar trend with local overdensity is reported
by Yan et al.(2008) for the DEEP2 Galaxy Redshift Survey at $z \sim
0.8$. An opposite trend with local density has been instead found by
Ma et al. (2008) in a cluster at $z=0.5$.  Our histogram shows that a
preference for underdense regions does not necessarily imply that
k+a's occur preferentially outside clusters: all our young k+a's
shown in Fig.~\ref{den2} reside in massive EDisCS
clusters. This trend with local density is compatible for example
with k+a's occurring at the edge of infalling structures, where they
first encounter the turbulent ICM. This scenario is also consistent
with the clustercentric distance distribution of k+a's shown below.

Interestingly, the k+a/active trend with density is paralleled by the
e(a)/active fraction that is consistent with a mild decrease with
density (right panel in Fig.~\ref{den2}). In contrast, the e(c)/active
fraction has the opposite trend, and sharply increases towards the
highest densities.

\begin{figure*}
\vspace{-4cm}
\centerline{\hspace{2cm}\includegraphics[width=14cm]{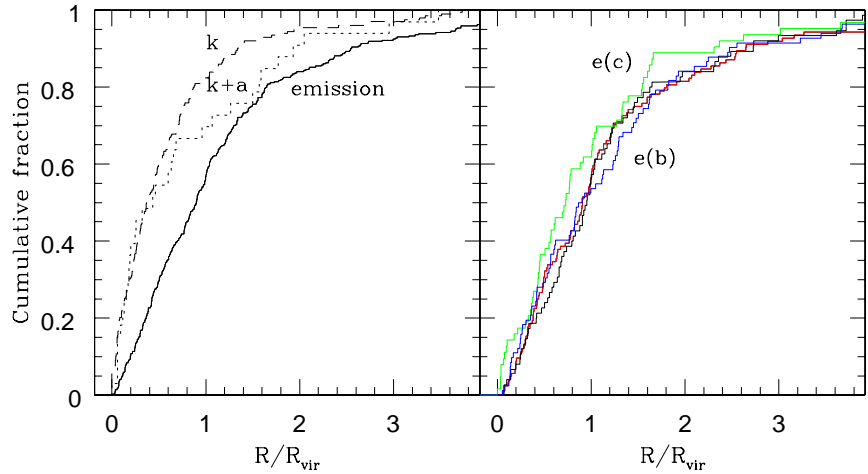}}
 \caption{Radial cumulative distribution of galaxies of different
 spectral types. {\bf Left.} K, k+a and all emission line galaxies. 
{\bf Right.} Emission line classes: e(a) (black), e(a)+ (red),
e(c) (green) and e(b) (blue).
 \label{cum}}
 \end{figure*}

Finally, we consider the distribution of clustercentric radial projected
distance of galaxies of different spectral types. Figure \ref{cum}
shows the cumulative radial distribution 
in units of virial radii. The virial radius
was computed as $R_{200}/1.14$ (Biviano et al. 2006
and Biviano 2007 private communication).
The distribution of all emission-line galaxies is shown in
the left panel and contrasted with the distributions of k and k+a
galaxies. The single emission-line classes are shown in the right
panel.

The star formation sequence corresponds to a progressive shift in
radial distributions: the oldest (k) galaxies are the most
centrally concentrated, the recently star-forming (k+a) are less
concentrated than k's and the currently star-forming (emission line
types) galaxies are the least concentrated, as found before e.g.  by
Dressler et al. (1999) and Moran et al. (2005). The distribution of
emission-line galaxies differs from that of k and k+a galaxies at the
99.9\% and 99.2\% level, respectively.  e(c)'s seem to be the most
centrally concentrated and e(b)'s the least concentrated among
emission-line galaxies, although numbers are too small for the
differences among the various emission types to be statistically
significant.

The radial distributions of galaxies within the cluster are consistent
with a scenario in which the longer a galaxy has been inside a cluster,
the longer ago was its star formation terminated and the longer is the
time it has had to dynamically relax within the cluster potential. The
projected local density distributions show a tendency for young k+a
galaxies to be located in low density cluster regions, as some of
their most likely progenitors: e(a) galaxies.

\section{Summary}

In this paper we have found that the incidence of k+a (post-starburst)
galaxies at intermediate redshifts depends strongly on environment.
K+a galaxies at $z=0.4-0.8$ preferentially reside in
clusters and, unexpectedly, in a subset of the groups, those with a 
low-[OII] fraction. In these environments, the star formation activity
is truncated on a short timescale ($<< 1$ Gyr) in a large fraction (20
to 30\%) of the recently 
star-forming population.

In constrast, there are proportionally fewer or even no k+a galaxies
in other environments, namely the field, the poor groups and the
groups with a high [OII] fraction.

These results are based on the observed variation of the fractions of
different spectral types with environment and are confirmed by the
study of cluster and group integrated composite spectra.

The quenching efficiency -- measured as the ratio between the number
of k+a galaxies and the number of galaxies with an ongoing or recent
star formation activity -- correlates with velocity dispersion.  More
massive systems have thus higher proportions of k+a galaxies, and higher
quenching efficiencies.  

In addition, low-[OII] systems, characterized by a low-[OII] fraction
for their mass, display an excess of k+a galaxies compared to other
systems of similar mass.  Moreover, all systems follow a tight
anticorrelation between the k+a/active galaxy fraction and the
fraction of [OII] emitters. This anticorrelation could imply that the
number of passively evolving galaxies in a cluster is closely linked
with the number of recently quenched galaxies.

Dusty starburst candidates present a very different environmental
dependence from post-starburst galaxies. They are numerous in all
environments at $z=0.4-0.8$, representing at least 20\% of the
star-forming population, but they are especially numerous among
star-forming galaxies in both low-[OII] and high-[OII] groups, where
they represent 45\% of the emission-line spectra (Table~6).  This
favors the hypothesis that a dusty starburst can be triggered by
mergers, that are expected to be common in groups.

Since dusty starburst candidates are present in similar proportions of
the star-forming population in high- and low-[OII] groups, while k+a
galaxies are present only in the latter type of groups, it is reasonable to conclude
that the star formation enhancement and the truncation of the star
formation are not necessarily associated phenomena and are caused by
different processes: the e(a) phase is not necessarily followed by a
k+a phase.

The spectral classification scheme, from passively evolving (k) to
post-starburst (k+a), to quiescently star-forming (e(c)) to starbursts
and starburst candidates (e(a), e(a)+ and e(b)) is also a sequence of
progressively later morphological types, lower galaxy stellar masses
and lower mass-to-light ratios.
We confirm that the properties of k+a galaxies are consistent with
them being observed in a transition phase, at the moment they are
rather massive S0s and Sa galaxies, presumably evolving from
star-forming later types into passively evolving early-type galaxies.
We also confirm that the clustercentric radial distribution is
consistent with this evolutionary connection between the different
types.

Interestingly, the observed frequency of k+a's 
shows no preference for the highest density regions. In fact, the
incidence of young k+a's suggests there is a higher k+a production rate
towards the low density regions within the cluster and group virial
radius, although the number of galaxies is too low to draw definite
conclusions.

\subsection{Discussion}

The most important and puzzling of our results are the high incidence
of k+a galaxies in some of the groups and the fact that the quenching
efficiency appears to increase with cluster velocity dispersion.

A possibility to be contemplated is that the velocity dispersions of
the [OII] outliers are severely underestimated.  This is rather
unlikely, for a number of reasons. First of all, they would have to be
extremely underestimated, and to be aligned with the rest of the
systems in the k+a fraction versus $\sigma$ relation, groups with
$\sigma \leq 200 \rm \, km \, s^{-1}$ should be clusters with $\sigma
\sim 800 \rm \, km \, s^{-1}$. The velocity dispersions of these
systems are based on a significant number of spectroscopically
confirmed members (17 in the case of Cl1119, 24 for Cl1420, for
example) and therefore have rather small formal errors. Moreover, a
sparse spectroscopic sampling usually leads to {\it overestimate}, not
underestimate, the $\sigma$. The weak lensing mass estimates do not
accomodate such large masses either (Clowe et al. 2006,
Milvang-Jensen et al. 2008).  A
severely underestimated mass is even harder to be explained in the
case of the two MORPHS outliers (crosses in Fig.~2), that
already have $\sigma =650-1000 \rm \, km \, s^{-1}$ and that to align
with the rest of the clusters should have $\sigma \geq 2000 \rm \, km
\, s^{-1}$. Overall, it is hard to see how significantly larger masses
could go undetected in the low-[OII] systems, although this is a
possibility that we cannot exclude for certain and awaits further
observational investigation.

Assuming that the measured velocity dispersions of the [OII] outliers are 
representative of their masses, we are left with two main questions:

a) what are the physical mechanisms that truncate star formation
in these systems and in the other environments? 

b) why do groups of similar masses divide in two subgroups having
respectively a quenching efficiency as high as those in clusters (the
low-[OII] groups) and a null quenching efficiency (the high-[OII
groups)?

In the following we attempt to delineate a working hypothesis, on the
basis of our and previous results. 

The fact that the quenching efficiency increases with cluster velocity
dispersion (Fig.~2) can be easily explained if the truncation is due
to the intracluster medium. Ram pressure stripping becomes more
efficient in more massive clusters and the density gradients at the
shock fronts of infalling structures should become steeper.  This
hypothesis is in agreement with previous evidence for a link between
k+a's and the hot intracluster medium in massive clusters (Poggianti
et al. 2004, Moran et al. 2007, Ma et al. 2008).

Theoretical calculations of ram pressure stripping through a static,
smooth intracluster medium find this to be effective only in the cores
of the most massive clusters. However, observations of stripped
galaxies (eg. Kenney et al. 2004, Vollmer et al. 2006) where stripping
is not expected to be efficient clearly challenge this simplistic
theoretical picture. Recent theoretical models have started to
investigate more realistic conditions in cluster outskirts, groups,
filaments and in the presence of galaxy tidal interactions, and to
assess the effects ot ICM turbulence, ICM subsctructure and shocks
during groups infall, finding that ram pressure can occur out to the
cluster virial radius (Tonnesen, Bryan \& van Gorkom 2007, Tonnesen \&
Bryan 2008, Kapferer et al. 2008).

At this point it is appropriate to wonder if an ICM origin can
accomodate the fact that bright k+a's are numerous in distant massive
clusters and rare in low-z massive clusters. The evolution of bright
cluster k+a's can be understood considering the evolution of the
infalling star-forming population. Clusters at $z=0$ with $\sigma >
500 \, \rm km \, s^{-1}$ have about 20\% of their galaxies that are
star-forming (Poggianti et al. 2006), as opposed to an average $\sim
50$\% in intermediate-z clusters (Table~4, see also Poggianti et
al. 2006 for the evolution of the star-forming fraction as a function
of cluster mass).  A quenching efficiency similar to that in distant
clusters (23\%, Table~6) would imply a global k+a fraction at $z=0$ of less
than 5\% ($0.20 \times 0.23 = 0.046$). A low k+a fraction at low-z can
therefore be explained as the result of a reduced ``fuel stream''
(rate of star-forming infalling galaxies), even if the quenching
efficiency remains similar to that in distant clusters as expected if
it is driven by the ICM. Such high quenching efficiency is also able
to account for the large number of faint k+a's in a cluster like Coma:
the ``fuel'' of star-forming galaxies is high at $z=0$ at the
magnitudes of dwarf galaxies, due to the downsizing effect.

We find the quenching efficiency to be comparable in clusters and
low-[OII] groups. Either the same physical mechanism or two different
mechanisms must be at work in these two environments.  For the
mechanism to be unique and be related to the intergalactic medium
(IGM), if one assumes that the trend of k+a's with $\sigma$ traces the
``standard'' trend of IGM with $\sigma$, the low-[OII] systems ought
to have an unusually dense medium for their mass.  In this case, the
major difference between low-[OII] and high-[OII] groups of similar
masses would be the presence/lack of this dense medium.

Indeed 3 out of 4 of the X-ray luminous groups with $\sigma \lesssim
400 \rm \, km \, s^{-1}$ observed at intermediate redshift by Jeltema
et al. (2007) have low-[OII] fractions that are
comparable to our outliers, but
1 out of the 4 exhibits a rather high [OII] fraction, intermediate
between that of our low- and high-[OII] groups. X-ray selected groups
were also found to have an overall low-[OII] content at $z=0.2-0.3$ by
Balogh et al. (2002).  On the other hand, the emission-line fractions
{\it inferred from the morphological mix} (and therefore highly
uncertain) in low-z groups detected in X-ray would qualify several of
these as probable high-[OII] systems (Zabludoff \& Mulchaey 1998,
Mulchaey et al. 2003). In the GEMS sample at low-z, Osmond \& Ponman
(2004) find that the fraction of spirals is anticorrelated with the
X-ray temperature with a large scatter, although they also find it
correlates with velocity dispersion. Therefore, these observations do
not provide an explanation for the different [OII] content at similar
$\sigma$ as we observe.

That different star formation properties in our two types of groups
are due to different IGM properties is in principle a viable
explanation, but entirely speculative based on our current
observations.  X-ray observations of large samples of groups with
different [OII] content are necessary to study the relation between
IGM properties and star formation.

Alternatively, the high quenching efficiency of low-[OII] groups could
be due to processes unrelated to the IGM. The most obvious candidate
mechanisms are tidal effects, due to tidal stripping, close galaxy
interactions or mergers. In this case, low-[OII] groups could be bound
objects in which tidal interactions are very effective, while
high-[OII] ``groups'' could be in fact systems of galaxies, for
example in filaments, that have not yet collapsed into a virialized
system. In this context, however, it is harder to understand why the
e(a)/active fraction is so similar in these two types of systems,
unless this fraction is unrelated to mergers. If the e(a)/active fraction
is related to mergers at some level, then it remains unclear why
the subsequent evolution involves a truncation of star formation only
in low-[OII] groups.  

Combinations of different processes can of course be at work: a
merger+starburst in all types of groups could largely exhaust the gas
and the latter could not be replenished only in those (presumably
low-[OII]) systems where the hot gas reservoir has been stripped by the
IGM, either because the density of the IGM is higher or because the
effects of the IGM are enhanced by the merging of different
subgroups. The IGM properties may also depend on the AGN activity 
history of the group, including powerful jets triggered 
in radio-loud quasars when two supermassive black holes
coalesce (Rawlings \& Jarvis 2004).

The only thing that appears clear is that
different physical (e.g. IGM or bound/unbound) conditions,
or a different growth history and dynamical status (recent subgroup merging)
must occur in groups with velocity dispersions in the range $150-400
\rm \, km \, s^{-1}$, rendering some of them efficient in suddenly
quenching star formation and some not. This may be connected with the
large scatter in star formation properties observed in groups
at low-z (Poggianti et al. 2006).

We now turn to the origin of k+a galaxies in the field and poor
groups. Observational evidence for a merger origin in the general
``field''\footnote{We stress again that the general field in other
studies comprises all non-cluster galaxies, and therefore includes
unknown percentages of our field, poor group and group populations.}
at intermediate redshifts, and a non-merger origin in clusters at
similar redshifts, was obtained by Tran et al. (2003) and
(2004). The evidence for mergers in distant field k+a's resembles the
merger signatures observed in local field k+a's (Zabludoff et
al. 1996, Yang et al. 2004, 2008).  A merger could, at least
temporarily, exhaust in a short time the available gas and thus give
rise to a k+a spectrum.  Within our dataset, k+a field galaxies with
HST imaging are too few for a statistical analysis of merger
occurrence. It is of course also possible that the field and poor
group galaxies that we observe are in fact in unidentified low-[OII]
groups.

Finally, following the discovery of galactic winds in Balmer-strong
galaxies at $z \sim 0.6$ in the SDSS (Tremonti et al. 2007), we have
looked for evidence for massive gas outflows from our spectra.  Some
of the processes discussed above could give rise to such a massive
outflow. For example, a strong starburst or feedback from a powerful
AGN phase, triggered by the merging of gas-rich galaxies, could
truncate star formation and produce a k+a spectrum.  As described in
detail in the Appendix below, we find no spectroscopic evidence for
such winds in the subset of galaxies with similar Balmer properties as
those in Tremonti et al. We note, however, that our conclusion is
limited to e(a) galaxies, because only one of our k+a spectra extends
to sufficiently blue wavelengths for this analysis, and therefore we cannot
exclude the presence of recent strong outflows in k+a galaxies.

\acknowledgments 
We would like to thank John Moustakas for helpful
discussions and for sharing unpublished results, Fabio Barazza for
providing his bar classification catalog, and the anonymous referee
whose report triggered several improvements in the paper.  BMP thanks
the Alexander von Humboldt Foundation and the Max Planck Instituut
f\"ur Extraterrestrische Physik in Garching for a very pleasant and
productive stay during which the work presented in this paper was
carried out. BMP acknowledges financial support from the INAF-National
Institute for Astrophysics through its PRIN-INAF2006 scheme. The Dark
Cosmology Centre is funded by the Danish National Research Foundation.



{\it Facilities:} \facility{VLT (FORS2)}, \facility{HST (ACS)}, \facility{NTT (SOFI)}.

\begin{table*}
\begin{center}
{\scriptsize
\caption{List of clusters.\label{tbl1}}
\begin{tabular}{llcclcccc}
\tableline\tableline
&&&&&&&& \\
Cluster name & Short name & $z$ & $N_{mem}$ & $\sigma$ $\pm{\delta}_{\sigma}$ & k+a\% & k+a/active &EW([OII]) & EW($H\delta$)\\
&&&& $\rm km \, s^{-1}$ && & \AA & \AA \\
\tableline
Clusters &&&&&&&&\\
 Cl\,1232.5-1250     &  Cl\,1232     & 0.5414  & 54&1080 $_{-89}^{+119}$  & 0.17$\pm$0.06 & 0.35$\pm$0.12 &1.0: & 1.5 \\ 
 Cl\,1216.8-1201     &  Cl\,1216     & 0.7943  & 67&1018 $_{-77}^{+73}$   & 0.13$\pm$0.04 & 0.23$\pm$0.08 &5.0  & 1.8 \\ 
 Cl\,1138.2-1133     &  Cl\,1138     & 0.4796  & 49& 732 $_{-76}^{+72}$   & 0.02$\pm$0.02 & 0.03$\pm$0.03 &14.3 & 2.1 \\  
 Cl\,1411.1-1148     &  Cl\,1411     & 0.5195  & 22& 710 $_{-133}^{+125}$ & 0.00$\pm$0.04 & 0.00$\pm$0.20 &4.5  & 1.5 \\ 
 Cl\,1301.7-1139     &  Cl\,1301     & 0.4828  & 35& 687 $_{-86}^{+81}$   & 0.09$\pm$0.05 & 0.14$\pm$0.08 &13.2 & 2.2 \\ 
 Cl\,1353.0-1137     &  Cl\,1353     & 0.5882  & 20& 666 $_{-139}^{+136}$  & 0.10$\pm$0.07 & 0.20$\pm$0.14 &8.9  & 1.3 \\ 
 Cl\,1354.2-1230     &  Cl\,1354     & 0.7620  & 21& 648 $_{-110}^{+105}$  & 0.14$\pm$0.08 & 0.19$\pm$0.11 &6.7  & 1.8 \\ 
 Cl\,1054.4-1146     &  Cl\,1054-11  & 0.6972  & 49& 589 $_{-70}^{+78}$   & 0.08$\pm$0.04 & 0.12$\pm$0.05 &6.6  & 1.5 \\ 
 Cl\,1227.9-1138     &  Cl\,1227     & 0.6357  & 22& 574 $_{-75}^{+72}$   & 0.00$\pm$0.04 & 0.00$\pm$0.07 &20.9 & 1.9 \\ 
 Cl\,1138.2-1133a    &  Cl\,1138a    & 0.4548  & 14& 542 $_{-71}^{+63}$  & 0.00$\pm$0.07 & 0.00$\pm$0.12 &2.3  & 1.6 \\ 
 Cl\,1202.7-1224     &  Cl\,1202     & 0.4240  & 19& 518 $_{-104}^{+92}$  & 0.00$\pm$0.05 & 0.00$\pm$0.17 &4.9  & 1.6 \\ 
 Cl\,1059.2-1253     &  Cl\,1059     & 0.4564  & 41& 510 $_{-56}^{+52}$   & 0.07$\pm$0.04 & 0.11$\pm$0.06 &6.3  & 1.7 \\ 
 Cl\,1054.7-1245     &  Cl\,1054-12  & 0.7498  & 36& 504 $_{-65}^{+113}$  & 0.09$\pm$0.05 & 0.16$\pm$0.09 &6.2  & 1.6 \\ 
 Cl\,1018.8-1211     &  Cl\,1018     & 0.4734  & 33& 486 $_{-63}^{+59}$   & 0.07$\pm$0.05 & 0.12$\pm$0.08 &9.9  & 1.6 \\ 
 Cl\,1227.9-1138a    &  Cl\,1227a    & 0.5826  & 11& 432 $_{-81}^{+225}$  & 0.08$\pm$0.08 & 0.09$\pm$0.09 &25.9 & 2.7  \\ 
 Cl\,1040.7-1155     &  Cl\,1040     & 0.7043  & 30& 418 $_{-46}^{+55}$   & 0.03$\pm$0.03 & 0.04$\pm$0.04 &13.4 & 2.2 \\ 
&&&&&&\\							      
Groups &&&&&&&& \\					      
High-OII &&&&&&&& \\
 Cl\,1037.9-1243     &  Cl\,1037     & 0.5783  & 16 &319 $_{-52}^{+53}$   & 0.00$\pm$0.06 & 0.00$\pm$0.06 &13.6 & 1.4: \\ 
 Cl\,1040.7-1155b    &  Cl\,1040b    & 0.7798  & 8 &259 $_{-52}^{+91}$   & 0.00$\pm$0.12 & 0.00$\pm$0.12 &17.6 & $>1.6$ \\ 
 Cl\,1103.7-1245b    &  Cl\,1103b    & 0.7031  & 11 &252 $_{-85}^{+65}$ & 0.00$\pm$0.09 & 0.00$\pm$0.09 &12.1 & 2.2 \\ 
 Cl\,1054.4-1146a    &  Cl\,1054-11a & 0.6130  & 8 &227 $_{-28}^{+72}$   & 0.00$\pm$0.12 & 0.00$\pm$0.12 &--   & -- \\ 
 Cl\,1040.7-1155a    &  Cl\,1040a    & 0.6316  & 11 &179 $_{-26}^{+40}$   & 0.00$\pm$0.09 & 0.00$\pm$0.09 &16.1 & 2.3 \\ 
Low-OII &&&&&&&& \\			       	  							
 Cl\,1301.7-1139a    &  Cl\,1301a    & 0.3969  & 17 &391 $_{-69}^{+63}$   & 0.13$\pm$0.09 & 0.29$\pm$0.21 &5.5  & 1.5 \\ 
 Cl\,1103.7-1245a    &  Cl\,1103a    & 0.6261  & 15 &336 $_{-40}^{+36}$   & 0.08$\pm$0.08 & 0.14$\pm$0.14 &6.1  & 2.0 \\ 
 Cl\,1420.3-1236     &  Cl\,1420     & 0.4962  & 24 &218 $_{-50}^{+43}$   & 0.13$\pm$0.07 & 0.27$\pm$0.16 &8.8  & 1.8 \\ 
 Cl\,1054.7-1245a    &  Cl\,1054-12a & 0.7305  & 10 &182 $_{-69}^{+58}$   & 0.11$\pm$0.11 & 0.20$\pm$0.20 &7.3  & 2.2 \\ 
 Cl\,1119.3-1129     &  Cl\,1119     & 0.5500  & 17 &166 $_{-29}^{+27}$   & 0.11$\pm$0.07 & 0.20$\pm$0.14 &6.9  & 2.1 \\
&&&&&&&&\\
Clusters + groups &&& 660 &&&&  9.8  & 1.8 \\
Clusters &&& 523 &&&&  9.8  & 1.7 \\
 Poor groups $z=0.4-0.8$ &&&  78 &&&& 12.4 & 2.4 \\
 Field $z=0.4-0.8$  &&& 123 &&&& 12.3 & 2.0 \\
Groups ``no excess'' &&& 71 &&&&  13.3  & 2.1 \\
``Excess'' clusters \& groups & & & 154 &&&&  6.3  & 1.9 \\
\tableline
\end{tabular}
}
\tablecomments{Col. (1): Cluster name. 
Col. (2): Short cluster name. 
Col. (3): Cluster redshift. Col. (4):
Number of spectroscopically confirmed members. Col. (5): Cluster velocity dispersion.
The numbers in columns (3), (4) and (5) are taken from Halliday et al. (2004), Milvang-Jensen et al. (2008) and Poggianti et al. (2006). Col. (6): Fraction of k+a galaxies, defined as number of
k+a galaxies/total number of members. Col (7): Fraction of k+a's among active galaxies,
defined as number of k+a galaxies/total number of active galaxies (see text in \S5).
Col. (8) and (9): Rest frame EW([OII]) and EW($\rm H\delta$)
of composite spectra (\S6). The Cl\,1054-11a composite spectrum is too noisy to yield reliable
line measurements. The definition of ``Groups ``no excess'''' and ``Excess'' clusters and groups is given
in \S6.
}
\end{center}
\end{table*}

\begin{table}
\begin{center}
{
\caption{Line strength in composite spectra\label{tbl2}}
\begin{tabular}{lcc}
\tableline\tableline
&& \\
Spec type & EW([OII]) & EW($H\delta$)\\
& \AA & \AA \\
k             & 0.0$^{+0.2}$  &  1.9$^{+0.1}_{-0.5}$ \\ 
k+a           & 0.9$^{+0.2}_{-0.2}$  &  3.0$^{+0.3}_{-0.2}$ \\
e(a)          & 15.1$^{+0.5}_{-2.0}$ &  $4.2$$^{+0.6}_{-0.4}$ \\
e(a)+         & 17.9$^{+0.2}_{-0.9}$ &  3.1$^{+1.1}_{-0.1}$ \\
e(c)          & 9.4$^{+0.6}_{-1.2}$  &  2.1$^{+0.2}_{-0.2}$ \\
e(b)          & 36.4$^{+1.0}_{-1.2}$ &  filled \\ 
\tableline
\tableline
\end{tabular}
}
\end{center}
\end{table}

\begin{table*}
\begin{center}
{\scriptsize
\caption{Raw spectral fractions.\label{tbl3}}
\begin{tabular}{lccccccccc}
\tableline\tableline
&&&&&&&&& \\
             & k(+k(e)) & k+a & e(a) & e(a)+ & e(c) & e(b) & e  &  $N_{?}$ & $N_{tot}$ \\ 
&&&&&&&&& \\
All clusters &   0.35$\pm$0.03(+0.07$\pm$0.01)  &  0.08$\pm$0.01  &  0.11$\pm$0.01  & 0.16$\pm$0.02  & 0.08$\pm$0.01  &  0.09$\pm$0.01  &  0.07$\pm$0.01   &  34 &  514\\

All groups   &   0.29$\pm$0.05(+0.08$\pm$0.02)  &  0.07$\pm$0.02  &  0.17$\pm$0.04  & 0.15$\pm$0.03  &  0.11$\pm$0.03 & 0.10$\pm$0.03  &   0.03$\pm$0.02   &   9 &  142\\

Groups low-OII&  0.48$\pm$0.08(+0.04$\pm$0.02)  & 0.12$\pm$0.04  &  0.09$\pm$0.03  & 0.08$\pm$0.03  &  0.10$\pm$0.04 &  0.05$\pm$0.03  &   0.04 $\pm$0.02  &   8 &  85 \\

Groups high-OII& 0.04$\pm$0.03(+0.13$\pm$0.05)  &  0.0$\pm$0.02  & 0.27$\pm$0.07  & 0.25$\pm$0.07 &  0.13$\pm$0.05 & 0.16$\pm$0.05  &   0.02$\pm$0.02   &   1 &  57 \\

Poor groups  &    0.07$\pm$0.03(+0.08$\pm$0.03)  &  0.03$\pm$0.02  &  0.14$\pm$0.04 &  0.31$\pm$0.07 &  0.06$\pm$0.03  &  0.08$\pm$0.03 &  0.24$\pm$0.06   &   6 &  78  \\

Field        &   0.13$\pm$0.03(+0.08$\pm$0.02)  &  0.04$\pm$0.02  &  0.10$\pm$0.03  & 0.22$\pm$ 0.04 &  0.06$\pm$0.02  & 0.22$\pm$0.04  &  0.14$\pm$0.03   &  11 &  140 \\ 

\tableline
\tableline
\end{tabular}
}
\tablecomments{Raw spectral fractions for the unweighted
magnitude-limited sample, with no radial constraint. The proportion of
k(in parenthesis, k(e)), k+a, e(a), e(a)+, e(c) and e(b) galaxies are
listed, together with the fraction of galaxies with at least a line in
emission but unknown $\rm H\delta$ strength (``e''), that cannot be
assigned to a specific emission-line class.  In the last two columns,
the number of spectra with an unknown spectral type (``?''), and the
total number of spectra. Errorbars are computed from Poissonian
statistics. 
The term ``All groups'' includes both low- and high-[OII]
groups, but {\it not} the poor groups.
}
\end{center}
\end{table*}

\begin{table*}
\begin{center}
{\scriptsize
\caption{Spectral fractions within $R_{200}$ weighted for incompleteness.\label{tbl4}}
\begin{tabular}{lccccccc}
\tableline\tableline
&&&&&&& \\
             & k(+k(e)) & k+a & e(a) & e(a)+ & e(c) & e(b) & e  \\
&&&&&&& \\
All clusters &     0.37$\pm$0.03(+0.06$\pm$0.01)  &  0.11$\pm$0.02     &  0.12$\pm$0.02  &  0.14$\pm$0.02  &  0.08$\pm$0.02  &  0.06$\pm$0.01  &   0.05$\pm$0.01 \\

All groups   &     0.31$\pm$0.07(+0.09$\pm$0.03)  &  0.05$\pm$0.03      &  0.24$\pm$0.05  & 0.07$\pm$0.03 &  0.14$\pm$0.04 &  0.06$\pm$0.02  &  0.04$\pm$0.02  \\

Groups low-OII&    0.54$\pm$0.10(+0.01$\pm$0.01)  & 0.10$\pm$0.04      &  0.15$\pm$0.05  & 0.01$\pm$0.01 &  0.15$\pm$0.05 &  0.0$\pm$0.02  &   0.03$\pm$0.02  \\

Groups high-OII&   0.03$\pm$0.03(+0.18$\pm$0.08)  &  0.0$\pm$0.02     &  0.33$\pm$0.11  & 0.14$\pm$0.07 &  0.13$\pm$0.07 & 0.13$\pm$0.04  &   0.05$\pm$0.03  \\

Poor groups  &      0.05$\pm$0.02(+0.08$\pm$0.04)  &  0.03$\pm$0.02      &  0.31$\pm$0.06  &  0.29$\pm$0.07 &  0.05$\pm$0.03  &  0.04$\pm$0.03  &  0.14$\pm$0.06   \\

Field        &     0.14$\pm$0.05(+0.08$\pm$0.04)  &  0.06$\pm$0.03      &  0.08$\pm$0.04  & 0.27$\pm$0.06  &  0.03$\pm$0.02  & 0.12$\pm$0.03  &  0.21$\pm$0.05   \\ 

\tableline
\tableline
\end{tabular}
}
\tablecomments{As Table~3, but for the magnitude-limited sample weighted
for spectroscopic incompleteness and considering only galaxies within $R_{200}$. The field and poor group samples have no radial limits.
The term ``All groups'' includes both low- and high-[OII]
groups, but {\it not} the poor groups.
}
\end{center}
\end{table*}

\begin{table*}
\begin{center}
{\scriptsize
\caption{Spectral fractions in the mass-selected sample.\label{tbl5}}
\begin{tabular}{lccccccccc}
\tableline\tableline
&&&&&&&&& \\
             & k(+k(e)) & k+a &  e(a) & e(a)+ & e(c) & e(b) & e  & $N_{?}$ &  $N_{tot}$ \\ 
&&&&&&&&& \\
All clusters &  0.44$\pm$0.04(+0.08$\pm$0.02) &  0.11$\pm$0.02  &  0.11$\pm$0.02 & 0.10$\pm$0.02 & 0.10$\pm$0.02 &  0.02$\pm$0.01  &  0.04$\pm$0.01 & 9 & 329 \\
All groups   &  0.32$\pm$0.07(+0.05$\pm$0.03) &  0.06$\pm$0.04  &  0.24$\pm$0.06 & 0.08$\pm$0.04 &0.20$\pm$0.05  &  0.01$\pm$0.01 & 0.03$\pm$0.01 & 2 & 69 \\
Groups low-OII&  0.60$\pm$0.13(+0.0$\pm$0.02)            &  0.12$\pm$0.05  &  0.12$\pm$0.05 & 0.01$\pm$0.01 &0.14$\pm$0.05  &   0.0$\pm$0.02  &   0.0$\pm$0.02 & 2 & 41   \\
Groups high-OII&  0.03$\pm$0.03(+0.10$\pm$0.07)& 0.0$\pm$0.04   &  0.37$\pm$0.12 & 0.15$\pm$0.07 &0.26$\pm$0.09  &  0.03$\pm$0.03 &  0.06$\pm$0.04 & 0 & 28 \\
Poor groups  &    0.09$\pm$0.05(+0.15$\pm$0.06)& 0.05$\pm$0.04  &  0.21$\pm$0.01 & 0.32$\pm$0.09 &0.09$\pm$0.04 & 0.03$\pm$0.03 & 0.06$\pm$0.04 & 1 & 40 \\
Field        &    0.20$\pm$0.08(+0.12$\pm$0.06)& 0.06$\pm$0.03  &  0.07$\pm$0.05 & 0.18$\pm$0.07 &0.04$\pm$0.03 & 0.11$\pm$0.03 & 0.21$\pm$0.05 & 4 & 47 \\

\tableline
\tableline
\end{tabular}
}
\tablecomments{As Table~3, but for the mass-selected sample weighted
for spectroscopic incompleteness and considering only galaxies
within $R_{200}$. The field and poor group samples have no radial limits.
The term ``All groups'' includes both low- and high-[OII]
groups, but {\it not} the poor groups.
}
\end{center}
\end{table*}

\begin{table}
\begin{center}
{
\caption{\label{tbl6}}
\begin{tabular}{lccc}
\tableline\tableline
&  k+a/active & e(a)/active  & e(a)/emission \\
Clusters        &  
0.23$\pm$0.04    &  0.23$\pm$0.04   &  0.33$\pm$0.05 \\

All groups      &  
0.09$\pm$0.05    & 0.39$\pm$0.10   &  0.46$\pm$0.12 \\  

Groups low-OII  &  
0.30$\pm$0.13   &  0.31$\pm$0.13  &  0.44$\pm$0.18 \\

Groups high-OII &   
0.00$\pm$0.02             &  0.43$\pm$0.14  &  0.46$\pm$0.15 \\

Poor groups     &  
0.06$\pm$0.05    &  0.28$\pm$0.09    &  0.32$\pm$0.11 \\ 

Field           &  
0.09$\pm$0.06    &  0.11$\pm$0.09    &  0.18$\pm$0.12 \\
\tableline
MORPHS clusters  &  
0.40$\pm$0.02   &  0.21$\pm$0.02    &  0.37$\pm$0.06 \\  

MORPHS ``field'' &  
0.09$\pm$ 0.03   &  0.15$\pm$0.04    &  0.20$\pm$0.07 \\
\tableline
\tableline
\end{tabular}
}
\tablecomments{Fractions in the mass-selected EDisCS sample within
$R_{200}$ weighted for spectroscopic incompleteness.  The ``active''
class includes: k+a, e(a), e(a)+, e(c), e(b), e.  ``Emission''
includes e(a), e(a)+, e(c) and e(b). The results for the
magnitude-limited sample are similar, always within the errorbar.
The term ``All groups'' includes both low- and high-[OII]
groups, but {\it not} the poor groups.
Also shown are results from the MORPHS collaboration, for their
approximately magnitude-limited sample.
}
\end{center}
\end{table}

 \begin{figure*}
\centerline{\hspace{-1.3cm}\includegraphics[width=16cm]{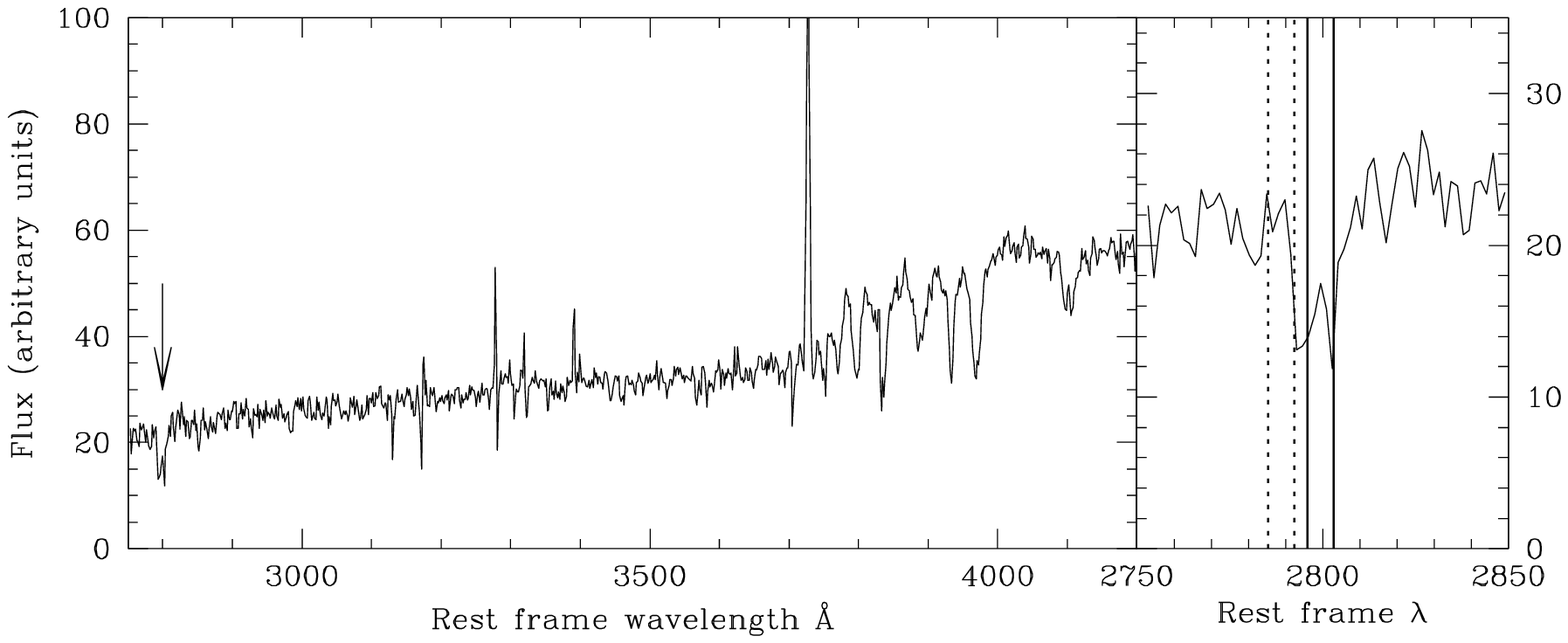}}
 \caption{Composite spectrum of our 6 highest S/N Balmer-strong 
spectra that cover the $Mg II \, 2796,2803$ lines.
{\bf Left} 
For this composite spectrum EW$(H\delta)= 4$ \AA $\,$ and EW(OII)=12
\AA. Strong Balmer absorption lines are clearly visible.
The arrow indicates the location of the $Mg II$ doublet.
{\bf Right} A zoom on the spectral region around the $Mg II$ lines.
Vertical solid lines indicate the expected line location
with no shift. The dotted lines show the shift corresponding
to the average velocity found by Tremonti et al. 2007,
(1140 $km \, s^{-1}$) at the average redshift of our spectra
$z=0.7$.
 \label{mg}}
 \end{figure*}

\appendix

\section{Massive outflows?}

Recently, Tremonti et al. (2007) have reported the discovery of
massive gas outflows in galaxies with strong Balmer lines in
absorption in SDSS spectra at $z\sim 0.6$.  Most of these spectra also
display an [OII] line in emission, judging from Fig.~1 in Tremonti et
al.  (2007).  Therefore, according to our classification scheme, they
would belong to our e(a) class.

Evidence for galactic winds in these galaxies comes from the $Mg II
2796,2803$ lines that are blueshifted by as much as $\sim 2000 km \,
s^{-1}$ with respect to the stars.  A large mass of outflowing gas can
be an efficient method for expelling the galactic gas and terminate
the star formation activity.  The end of the star formation in this
case would be related to a phenomenom intrinsic to the galaxy itself
such as an AGN.

184 of our spectra reach sufficiently out in the blue to cover the $Mg
II 2796,2803$ lines. Of these, 31 are cluster members, and of these 13
have spectral characteristics similar to those of Tremonti et al.,
i.e. have strong Balmer lines in absorption and relatively weak [OII]
lines. The great majority of these galaxies belong to our e(a)
spectral class, with the exception of 1 k+a.  The resemblance of these
spectra with those of the galaxies in Tremonti et al. sample can be
appreciated comparing our Fig.\ref{mg} with their Fig.~1.

Measuring the position and strength of the Mg II lines in the
spectra of these 13 galaxies, we are able to identify the Mg
II lines in most cases and we do not detect significant wavelength
shifts in any case.  Fig.\ref{mg} shows the composite spectrum of our
6 best S/N cases, with the strong Balmer absorption lines
(EW$(H\delta)$= 4 \AA), the relatively weak [OII] (EW=12 \AA) and the
Mg II lines in place at approximately their rest frame wavelength
(right panel in Fig.\ref{mg}).

Moreover, Tremonti et al. (2007) have shown that there is an observed
correlation between the outflow velocity $v$ and the galaxy absolute B
luminosity in both low-z and high-z samples. Given the moderate
typical B luminosity of our small sample ($<M_B> = -20.4$), the lack
of a blueshift in the Mg II line is consistent with the $v-M_B$
correlation.

Thus, we find that galaxies with strong Balmer absorption and weak to
moderate emission in our sample do not display signs of galactic
winds. Our result shows that this combination of optical spectral
features does not need to be generally associated with AGN-powered
outflows. The Tremonti et al. galaxies were targeted for SDSS
spectroscopy as quasar candidates, though they were subsequently
classified as galaxies. Selection criteria other than the strength of
$\rm H\delta$ and [OII] must play a role for identifying galaxies with
galactic winds.

\end{document}